\documentclass{article}
\usepackage{arxiv}
\usepackage[utf8]{inputenc}
\usepackage[T1]{fontenc}
\usepackage{lmodern}          
\usepackage{amsmath,amssymb,amsfonts}
\usepackage{amsthm}            
\usepackage{mathtools}         
\usepackage{bm}                
\usepackage{graphicx}          
\usepackage{caption}
\usepackage{subcaption}        
\usepackage{booktabs}          
\usepackage{array}
\usepackage{enumitem}          
\usepackage{geometry}          
\usepackage{hyperref}          
\hypersetup{
    colorlinks=true,
    linkcolor=blue,
    citecolor=blue,
    urlcolor=blue
}
\usepackage[numbers,sort&compress]{natbib} 
\usepackage{doi}               
\usepackage{algorithm}
\usepackage{algpseudocode}     
\usepackage[table]{xcolor}

\definecolor{softgreen}{RGB}{220,240,220}
\definecolor{softyellow}{RGB}{255,245,200}
\definecolor{paleCinnabar}{RGB}{242,200,195}
\definecolor{celadonGreen}{RGB}{205,225,215}
\definecolor{classicCeladon}{RGB}{210,220,215}

\newcommand{\highlightgreen}[1]{\cellcolor{classicCeladon!75}#1}

\theoremstyle{definition}

\theoremstyle{remark}

\newcommand{\sout}[1]{}

\usepackage{titling}

\title{ATLAS-NN: \underline{A}daptive \underline{T}ransfer \underline{L}earn\underline{a}ble \underline{S}ymplectic-aware Neural Network for Long-Time Hamiltonian Dynamics 
}

\usepackage[affil-it]{authblk}

\author[1]{Changhong Mou\thanks{These authors contributed equally to this work.}}
\author[2,3]{Dinghua Xu\protect\footnotemark[1]}
\author[2]{Xiyue Zuo}
\author[2]{Keji Liu}
\author[2]{Yeyu Zhang\thanks{Corresponding author: \href{mailto:zhangyeyu@mail.shufe.edu.cn}{zhangyeyu@mail.shufe.edu.cn}}}
\affil[1]{Department of Mathematics and Statistics, Utah State University, Logan, UT, USA}
\affil[2]{School of Mathematics, Shanghai University of Finance and Economics, Shanghai, China}
\affil[3]{Department of Mathematical Sciences, School of Science, Zhejiang Sci-Tech University, Hangzhou, China}
\affil[ ]{\small \textit{E-mail addresses:}
\href{mailto:changhong.mou@usu.edu}{changhong.mou@usu.edu} (Changhong Mou),
\href{mailto:dhxu6708@mail.shufe.edu.cn}{dhxu6708@mail.shufe.edu.cn} (Dinghua Xu),
\href{mailto:zuo_xy821@163.com}{zuo\_xy821@163.com} (Xiyue Zuo),
\href{mailto:liu.keji@mail.shufe.edu.cn}{liu.keji@mail.shufe.edu.cn} (Keji Liu),
\href{mailto:zhangyeyu@mail.shufe.edu.cn}{zhangyeyu@mail.shufe.edu.cn} (Yeyu Zhang).}

\makeatletter
\def\@fnsymbol#1{\ensuremath{\ifcase#1\or *\or \dagger\or \ddagger\or
   \mathsection\or \mathparagraph\or \|\or **\or \dagger\dagger
   \or \ddagger\ddagger \else\@ctrerr\fi}}
\makeatother
\date{\today}

\begin{document}

\maketitle

\begin{abstract}
Modeling Hamiltonian systems over long temporal intervals remains a significant challenge due to intrinsic multiscale structures and rapid nonlinear transitions. While Hamiltonian Neural Networks (HNNs) incorporate geometric invariants to improve stability, they typically rely on a fixed, externally prescribed temporal structure. This lack of adaptability often leads to accumulated phase errors and degraded accuracy in systems with heterogeneous temporal scales.
To address these limitations, we put forward the Adaptive Transfer Learnable Symplectic-aware Neural Network (ATLAS-NN). Our framework augments the HNN architecture with a learnable temporal scaling mechanism that parametrize a nonlinear mapping of time, automatically adapting to the system's intrinsic complexity. We propose a two-stage transfer learning strategy: the model is first trained on a short-time \textit{source} interval to identify the Hamiltonian structure and optimal temporal reparameterization; the learned scaling function is then frozen and transferred to an extended \textit{target} interval for fine-tuning. Numerical experiments on nonlinear oscillators and the chaotic H\'enon--Heiles system demonstrate that ATLAS-NN provides a more efficient alternative to standard HNNs and traditional symplectic integrators, yielding nearly an order of magnitude reduction in long-time prediction error.
\end{abstract}



\section{Introduction}
Scientific machine learning (SciML) has rapidly become a surrogate framework for solving differential equations arising in fluid dynamics, materials science, plasma physics, and many multi-physics systems \cite{carleo2019machine,cai2021physics,lu2021learning,lifourier,mou2023combining,thiyagalingam2022scientific}. 
By integrating data-driven models with the underlying physical laws, modern neural network architectures, such as physics-informed neural networks (PINNs), neural operators, and neural ODE methods, have achieved notable success in approximating complex dynamical behavior and accelerating numerical simulation \cite{cai2021physics,lifourier,rackauckas2020universal,chen2018neural,lu2021learning,ruiz2023neural}. 
However, many real-world dynamical systems remain difficult to model due to their intrinsic multiscale structure, rapidly changing temporal dynamics, or strong nonlinear couplings. Effectively simulating such phenomena often requires extremely fine temporal resolution. State of the art SciML methods may struggle to represent such dynamics when the system evolves on heterogeneous time scales. 
For a broad class of PDE- and ODE-driven physical systems, the evolution can be expressed in a Hamiltonian framework \cite{koopman1931hamiltonian,abraham1978foundations,marsden2013introduction,goldstein2011classical,sharma2022hamiltonian},
in which the dynamics preserve key geometric structures,
notably known as symplecticity, energy, and phase-volume invariance. 
These invariants fundamentally determine the long-term qualitative behavior of the solution, and putting them has long been central to geometric numerical integration. A key feature of Hamiltonian dynamics is that the preservation of these geometric structures directly governs long-time stability and the quantitative accuracy of the trajectories.
Classical geometric integrators, such as symplectic Euler \cite{donnelly2005symplectic,leimkuhler2004simulating,hairer2006structure,feng2006symplectic} and variational Runge–Kutta methods \cite{bottasso1997new}, are proposed to exploit these invariants and are widely used for long-time simulations. However, their accuracy and efficiency may deteriorate when the dynamics involve multiscale temporal behavior, rapid frequency changes, or localized fast variations. In such regimes, fixed-step integrators tend to either accumulate significant numerical error or require prohibitively small timesteps, making long-time simulation computationally expensive.

A notable development along this direction is the Hamiltonian Neural Network (HNN) frameworks which has been formulated in two different ways by Greydanus et al.~\cite{greydanus2019hamiltonian}
and by Mattheakis et al.~\cite{mattheakis2022hamiltonian}. Although both approaches exploit Hamilton's equations to incorporate geometric structure into neural network frameworks, they differ fundamentally in how this structure is enforced and in the learning setup. As a result, the two formulations provides different advantages and
trade-offs.
In particular, the data-driven HNN in \cite{greydanus2019hamiltonian} learns a parametric Hamiltonian $H_\theta$ directly from observations of trajectories, which makes it broadly applicable when the governing equations are unknown or only partially trusted; it can also generalize across initial conditions once $H_\theta$ is identified. 
However, its performance depends strongly on the quantity and quality of data, and limited coverage of the relevant state-space regions can prevent generalization and lead to unreliable extrapolation, particularly for long-time predictions.
By contrast, the symplectic-aware approach (symplectic-aware HNN) in \cite{mattheakis2022hamiltonian}, similar in spirit to physics-informed neural networks (PINNs) \cite{cai2021physics,raissi2019physics}, requires no training data and instead enforces Hamilton's equations (and optionally energy conservation) through a residual loss, which can yield physically consistent trajectories even in data-insufficient settings. 
Its main limitation is that it presumes accurate knowledge of the governing equations and typically produces a solution representation tied to the particular system/initial condition being trained, so it may require retraining for new parameter regimes or initial conditions.
 In addition, as in PINNs, the training can become difficult, especially over long time intervals: the solution may exhibit multi-temporal scale features or increasingly complex behavior, and the optimization may
struggle to enforce the physics constraints uniformly across the entire time window.

The symplectic-aware HNN learns a Hamiltonian function whose gradients generate the equations of motion, which enables the model to produce trajectories that preserve key invariants. Importantly, this formulation requires no ground-truth trajectory data; the optimization depends solely on satisfying Hamilton’s equations. Prior studies demonstrate that HNNs can accurately capture nonlinear oscillators and chaotic Hamiltonian systems, often requiring fewer evaluation points than traditional symplectic integrators to achieve comparable accuracy. 
However, plain vanilla HNNs share a critical limitation: the temporal structure of the dynamics is fixed and externally prescribed \cite{mattheakis2022hamiltonian,desai2021port}. Although some variants introduce an additional parametrized function of time to increase expressiveness or account for external forcing, this time-dependent component remains fixed in form and does not allow the network to adapt the notion of time itself. As a result, HNNs struggle when the underlying dynamics evolve on heterogeneous temporal scales, such as in stiff Hamiltonian flows, fast–slow interactions, or systems exhibiting intermittent rapid transitions. 
Without the ability to learn a time reparameterization, the model cannot allocate resolution where the dynamics demand it, often leading to accumulated phase errors or degraded long-time accuracy. 
To address this limitation, we put forward the \underline{A}daptive \underline{T}ransfer \underline{L}earn\underline{a}ble \underline{S}ymplectic-aware Neural Network (ATLAS-NN), a novel architecture that augments the Hamiltonian learning framework with a  data-driven temporal scaling mechanism. Instead of relying on a fixed $f(t)$ or externally prescribed temporal behavior, ATLAS-NN learns a nonlinear mapping of time that automatically stretches or compresses the temporal domain according to the intrinsic complexity of the system. Moreover, the proposed ATLAS-NN enables transfer learning task (cf. the PINN setting in \cite{pellegrin2022transfer}) for Hamiltonian systems by pretraining on short-time trajectories and then fine-tuning to achieve accurate long-time prediction.
In particular, we first train ATLAS-NN on a \textit{short time interval} to learn the Hamiltonian structure together with a \textit{temporal scaling (reparameterization) function} $f(t)$ that captures local time scale variations over the training window. We then \textit{freeze the learned $f(t)$} and fine-tune only the remaining network components on an extended \textit{long time interval}, so the model retains its learned time adaptation while efficiently adjusting its dynamics representation for accurate long time prediction.

The rest of the paper is organized as follows. Section \ref{sec:atlas-net} reviews the general Hamiltonian neural network (HNN) and introduces the proposed ATLAS-NN. Section \ref{sec:results} presents numerical experiments demonstrating ATLAS-NN in a transfer-learning setting, including the short time prediction and both the source and target tasks in the transfer learning task. Finally, Section \ref{sec:conclusion} concludes the paper and discusses directions for future work.


\section{Adaptive Transfer Learnable Symplectic-aware Neural Network \label{sec:atlas-net}}

\subsection{Hamiltonian System \label{ss:ham}}
Originally introduced as a general reformulation of classical mechanics, Hamiltonian mechanics provides a different perspective for describing the evolution of dynamical systems on a symplectic manifold. While initially developed for classical systems, its formalism has become ubiquitous in theoretical physics, underpinning fields ranging from statistical mechanics to quantum field theory~\cite{wiggins2003introduction,goldstein1950classical, sakurai2020modern}.

Consider a system described by the phase space coordinates or the so-called canonical coordinates: $\mathbf{z} = (\mathbf{q}, \mathbf{p})^\top \in \mathbb{R}^{2d}$, where $\mathbf{q} \in \mathbb{R}^d$ represents the generalized coordinates and $\mathbf{p} \in \mathbb{R}^d$ denotes the conjugate momenta. The dynamics of the system are described by a scalar function $\mathcal{H}: \mathbb{R}^{2d} \to \mathbb{R}$, known as the Hamiltonian, which typically represents the total energy. The time evolution of the state $\mathbf{z}$ is determined by Hamilton's equations:
\begin{equation}
    \frac{d\mathbf{z}}{dt} = \mathbf{J} \nabla \mathcal{H}(\mathbf{z}), \quad \text{where} \quad \mathbf{J} = 
    \begin{pmatrix} 
    \mathbf{0}_d & \mathbf{I}_d \\ 
    -\mathbf{I}_d & \mathbf{0}_d 
    \end{pmatrix}.
    \label{eqn:hamiltonian_dynamics}
\end{equation}
Here, $\mathbf{J}$ is the $2d \times 2d$ skew-symmetric symplectic matrix, $\mathbf{I}_d$ is the $d \times d$ identity matrix, and $\nabla \mathcal{H} = (\nabla_{\mathbf{q}}\mathcal{H}, \nabla_{\mathbf{p}}\mathcal{H})^\top$ is the gradient of the Hamiltonian with respect to $\mathbf{z}$.
Equation~\eqref{eqn:hamiltonian_dynamics} defines a Hamiltonian vector field, $\mathbf{F}_{\mathcal{H}}(\mathbf{z}) = \mathbf{J} \nabla \mathcal{H}(\mathbf{z})$. Unlike the gradient field $\nabla \mathcal{H}$, which points in the direction of steepest ascent, the symplectic gradient $\mathbf{J} \nabla \mathcal{H}$ generates a flow that is orthogonal to the gradient. Consequently, the Hamiltonian is a conserved quantity along the flow, a property derived directly from the skew-symmetry of $\mathbf{J}$:
\begin{equation}
    \frac{d\mathcal{H}}{dt} = (\nabla \mathcal{H})^\top \frac{d\mathbf{z}}{dt} = (\nabla \mathcal{H})^\top \mathbf{J} (\nabla \mathcal{H}) = 0.
    \label{eqn:energy_conservation}
\end{equation}
This structure-preserving property makes Hamiltonian mechanics particularly powerful for modeling complex, high-dimensional systems where energy conservation is paramount, such as $N$-body celestial mechanics, fluid dynamics, and quantum many-body systems~\cite{reichl2016modern, zagoskin1998quantum}.

As analytical solutions are rarely available for such systems, numerous numerical solvers have been proposed over the past decades. Standard integrators, such as the explicit Runge-Kutta methods, generally fail to preserve the geometric properties of the flow, leading to correct short-term accuracy but significant long-term energy drift. Consequently, Hamiltonian systems are typically solved using \textit{symplectic integrators}, which are designed to preserve the symplectic 2-form of the phase space.
A canonical first-order example is the Symplectic Euler (SE) method. For a separable Hamiltonian $\mathcal{H} = T(\mathbf{p}) + V(\mathbf{q})$, the state updates over a time step $\Delta t$ follows:
\begin{equation}
    q_i^{(n+1)} = q_i^{(n)} + \Delta t \frac{\partial T}{\partial p_i^{(n)}}, \qquad 
    p_i^{(n+1)} = p_i^{(n)} - \Delta t \frac{\partial V}{\partial q_i^{(n+1)}}.
    \label{eq:symplectic_euler}
\end{equation}
While this scheme preserves the symplectic structure of the discrete map $\mathbf{z}^{(n)} \mapsto \mathbf{z}^{(n+1)}$, it does not conserve the exact Hamiltonian $\mathcal{H}$. Instead, it exactly conserves a \textit{modified} or \textit{shadow} Hamiltonian $\tilde{\mathcal{H}}$, expressed as a power series expansion in $\Delta t$. This property ensures that the computed energy oscillates within a bounded interval of $\mathcal{O}(\Delta t)$ rather than drifting linearly or exponentially, making such methods essential for long-time simulations.

\subsection{ Symplectic-aware Hamiltonian Neural Network}

Mattheakis et al.~\cite{mattheakis2022hamiltonian} introduced a Hamiltonian Neural Network (HNN) to solve the dynamical systems described in Section~\ref{ss:ham}. HNN employs a parametric neural representation to approximate continuous trajectories. Rather than learning the discrete map directly, this framework approximates the flow map $\mathbf{z}(t)$ via the ansatz:
\begin{align}
    \hat{\mathbf{z}}(t) = \mathbf{z}(0) + f(t)\,{\bf N}(t; \theta), 
    \label{eq:param_sol_ref}
\end{align}
where ${\bf N}(t; \theta) \in \mathbb{R}^{2d}$ denotes a multi-output feed-forward neural network, and $f(t)$ is a scalar temporal factor that enforces the initial condition exactly via $f(0)=0$.
In this formulation, the network parameters $\theta$ are determined by minimizing the residual of the symplectic dynamics evaluated at a set of discrete sampled times $\{t_n\}_{n=1}^K \subset [0,T]$:
\begin{align}
    \mathcal{L} = \frac{1}{K}\sum_{n=1}^K 
    \Big\|
    \dot{\hat{\mathbf{z}}}(t_n) - \mathbf{J}\,\nabla_{\hat{\mathbf{z}}}\mathcal{H}\!\big(\hat{\mathbf{z}}(t_n)\big)
    \Big\|_2^2 + \lambda_{reg} \mathcal{L}_{\mathrm{reg}},
    \label{eq:hnn_loss_ref}
\end{align}
where $\lambda_{reg}$ is a regularization parameter.
Time derivatives $\dot{\hat{\mathbf{z}}}$ and gradients $\nabla\mathcal{H}$ are computed via automatic differentiation. To improve stability over long integration horizons, \cite{mattheakis2022hamiltonian} includes an optional energy regularization term:
\begin{align}
    \mathcal{L}_{\mathrm{reg}} = \frac{1}{K}\sum_{n=1}^K \Big( \mathcal{H}(\hat{\mathbf{z}}(t_n)) - E_0 \Big)^2, \quad \text{with} \quad E_0 = \mathcal{H}(\mathbf{z}(0)),
\end{align}
where $\lambda_{reg}$ is regularization coefficient. 
The methodology suggests that introducing mild random jitter $t_n \mapsto t_n + \varepsilon$ per epoch improves the generalization of the solution.
The choice of the temporal scaling factor $f(t)$ is noted as critical for scaling. While a linear choice $f(t)=t$ satisfies the initial condition, it grows unbounded. The referenced work therefore utilizes a bounded alternative,
\begin{align}
    f(t) = 1 - e^{-t},
\end{align}
to maintain a well-scaled parameterization over certain time interval. 


\subsection{Adaptive Transfer Learnable Mechanism\label{ss:adaptive}}

Standard parametric solvers often struggle with stiff Hamiltonian systems characterized by diverse temporal scales that ranges from rapid transients to slowly varying quasi-periodic drifts. To mitigate this, we introduce a learnable temporal scaling mechanism that decouples the training time interval from the physical time-scale of the dynamics. Crucially, this mechanism facilitates \textit{temporal transfer learning} \cite{weiss2016survey,torrey2010transfer,zhuang2020comprehensive}, allowing the model to generalize representations learned from short-term trajectories to effectively adapt to the prediction of long-term dynamical evolution.

\subsubsection{Learnable temporal scaling mechanism }
To illustrate the learnable temporal scaling mechanism, we define a class of admissible scaling functions $f(t; \gamma): \mathbb{R}_{\ge 0} \to [0, 1)$ parameterized by $\gamma$, subject to the boundary condition $f(0; \gamma) = 0$ and the strict monotonicity constraint $\dot{f}(t; \gamma) > 0$. The ansatz for the trajectory is defined as:
\begin{align}
    \hat{\mathbf{z}}(t) = \mathbf{z}_0 + f(t; \gamma)\, \mathbf{N}(t; \theta).
    \label{eq:ansatz_adaptive}
\end{align}
The temporal derivative, which must satisfy the symplectic flow condition, is obtained via the chain rule:
\begin{align}
    \dot{\hat{\mathbf{z}}}(t) = \underbrace{\dot{f}(t; \gamma)\,\mathbf{N}(t; \theta)}_{\text{Transient scaling}} + \underbrace{f(t; \gamma)\,\dot{\mathbf{N}}(t; \theta)}_{\text{Asymptotic dynamics}}.
    \label{eq:adaptive_derivative}
\end{align}
Equation \eqref{eq:adaptive_derivative} requires that $\dot{f}$ acts as an adaptive mollifier function. Initially, when $\dot{f}$ is large, the gradient flow prioritizes satisfying the initial velocity constraints; as $t \to \infty$ and $\dot{f} \to 0$, the dynamics become fully governed by the neural network $\mathbf{N}$.
In particular, we propose two different parametric schemes for $f(t; \gamma)$ that allows the framework to adapt to the specific temporal profile of the underlying physics.
\paragraph{Scheme I: $\tanh$ function.} 
We consider a hyperbolic tangent profile:
\begin{align}
    f(t; \gamma) = \tanh(m t), \qquad \gamma = \{m\},
    \label{eq:ft_tanh}
\end{align}
where $m > 0$ determines the characteristic transition rate. This form ensures a bounded mapping with linear behavior near $t=0$ and symmetric exponential approach to the asymptotic regime ($f \to 1$).

\paragraph{Scheme II: Exponential function.} 
To capture more complex, asymmetric transient behaviors, we utilize a generalized rational exponential family:
\begin{align}
    f(t; \gamma) = \frac{1 - e^{-\alpha t}}{1 + \beta e^{-\alpha t}}, \qquad \gamma = \{\alpha, \beta\}.
    \label{eq:ft_frac2}
\end{align}
Here, $\alpha \in \mathbb{R}^+$ represents the inverse characteristic time-scale, while $\beta \in \mathbb{R}_{\ge 0}$ controls the curvature of the transition. This scheme offers strictly richer representational capacity: in the limit $\beta \to 0$, it recovers standard exponential decay, while for $\beta=1$, it reproduces the sigmoidal profile of the hyperbolic tangent (specifically $\tanh(\frac{\alpha t}{2})$). By learning $\gamma$ jointly with the network parameters $\theta$, the framework automatically discovers the intrinsic time-scale of the system.

\begin{figure}[H]
    \centering
    \includegraphics[width=.8\linewidth]{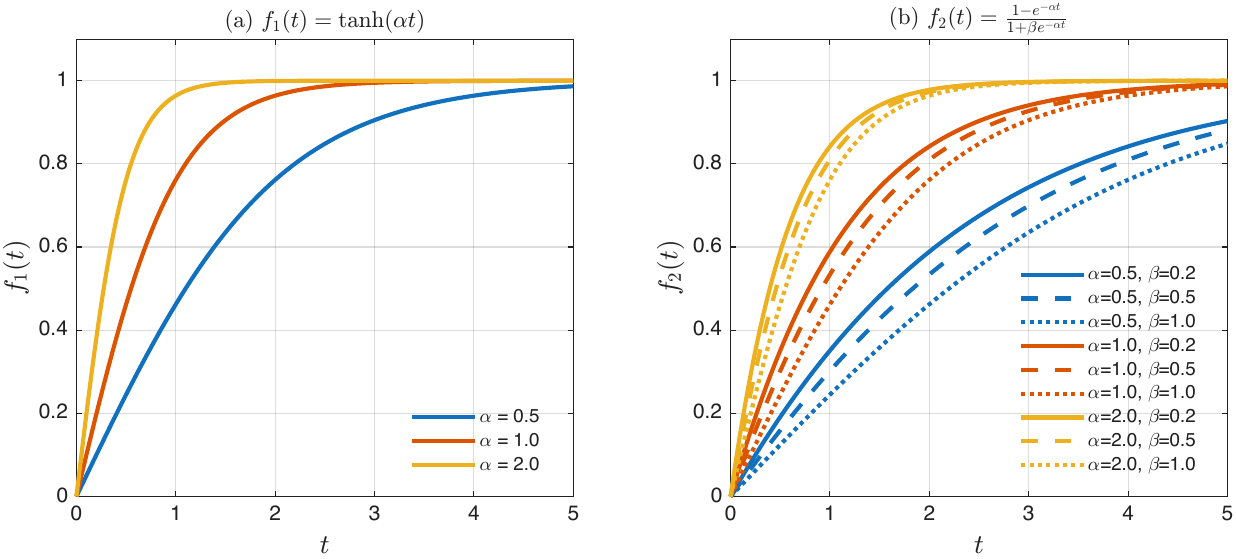}
\caption{Behavior of the adaptive temporal scaling function $f(t; \gamma)$ with different parameter values: (left). $f(t; \gamma) = \tanh(mt)$, (right). $f(t; \gamma) = \frac{1 - e^{-\alpha t}}{1 + \beta e^{-\alpha t}}$.}    \label{fig:f-fun}
\end{figure}

\subsubsection{Transfer learning for long time dynamics}
\begin{figure}[H]
    \centering
    \includegraphics[width=0.8\linewidth]{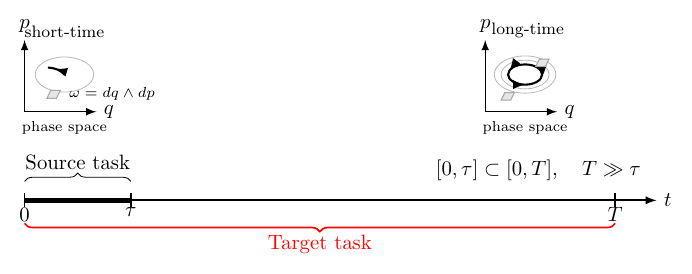}
\caption{Illustration of the transfer learning strategy used to extend the learned dynamics from short-time intervals to long-time intervals.}    \label{fig:transfer-learning}
\end{figure}
Simulating Hamiltonian systems inherently requires accuracy over long time domain, as the physical validity of a model is determined by its ability to conserve invariants, such as total energy, indefinitely. A neural network that approximates the vector field accurately only for short intervals may still suffer from accumulated errors that destroy the system's geometric properties over time. However, training a neural network directly on long-term trajectories is computationally prohibitive and susceptible to optimization instability, primarily due to the vanishing or exploding gradients that may arise when backpropagating through a numerical integrator over many time steps.

The proposed adaptive transfer learnable mechanism in Section \ref{ss:adaptive} can be easily used to incorporate transfer learning strategies to overcome these challenges. Transfer learning is a machine learning technique where a model developed for a source task is reused as the starting point for a model on a related target task. In the present context, we define the source task as learning the dynamics over short time intervals that can efficiently and effectively capturing the temporal structure of the Hamiltonian system; this is also illustrated in Figure \ref{fig:transfer-learning}. These pre-trained parameters are then transferred to initialize the training for the target task: long time trajectory.
In particular, utilizing the ansatz in equation \eqref{eq:ansatz_adaptive}, we take a selective parameter update strategy. During the short time source task, the time-parameterization coefficient $\gamma$ is held fixed at the value converged. Conversely, the neural network weights $\theta$ remain trainable. This strategy preserves the identified characteristic time scale that allows the network to fine-tune the Hamitonian system representation to preserve the symplectic form over the long time interval.

\begin{figure}[H]
    \centering
    \includegraphics[width=\linewidth]{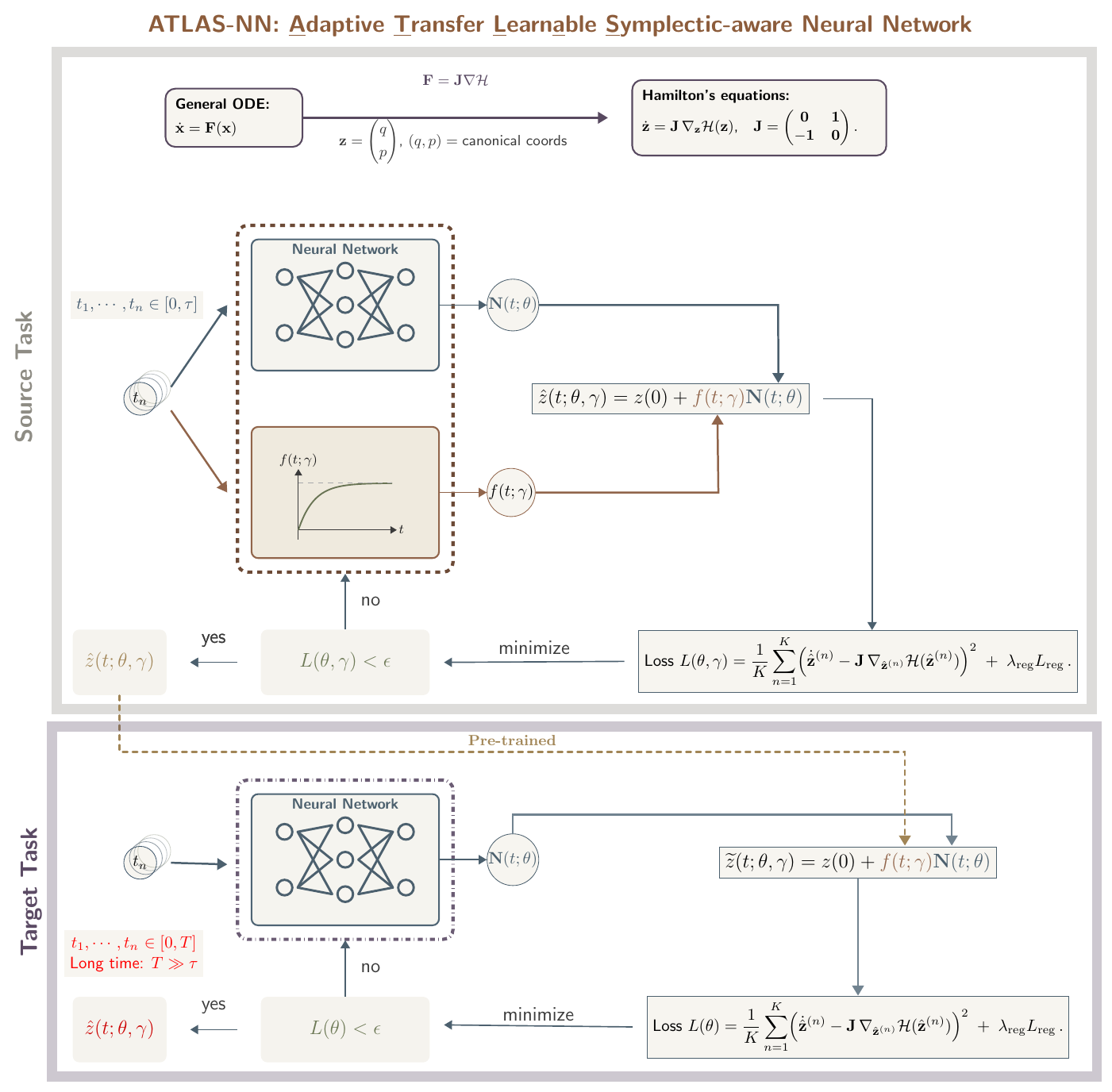}
    \caption{Illustration of ATLAS-NN: \underline{A}daptive \underline{T}ransfer \underline{L}earn\underline{a}ble \underline{S}ymplectic-aware Neural Network. The top panel illustrates the source task conducted over a short-time dynamics, where a neural network is pre-trained to approximate the Hamiltonian system defined by canonical coordinates. This initial phase is governed by a physics-informed loss function that penalizes deviations from the symplectic gradient flow and incorporates structural regularization to ensure the learned dynamics respect the underlying physical manifold. The bottom panel depicts the target task, where the framework transitions to a long-time period that significantly exceeds the original training window. In this transfer-learnable stage, the pre-trained model is combined with a learnable, time-parameterized scaling function to adapt the solution for extended temporal stability.}
    \label{fig:hnn}
\end{figure}

\subsection{Summary of ATLAS-NN}
The architecture and key features of the ATLAS-NN are illustrated in Figure \ref{fig:hnn}. 
The ATLAS-NN addresses the limitations of standard neural solvers in Hamiltonian systems by decoupling the initial training task from the long-term predictive task. The framework begins with a source task conducted over a short-time interval, during which the neural network learns to approximate the Hamiltonian dynamics in canonical coordinates. This stage is optimized via a physics-informed loss function that penalizes deviations from the symplectic gradient flow and incorporates energy regularization to ensure the learned dynamics respect the conservation laws of the underlying physical manifold.
Once the source task is finished, the framework transitions to the target task, which focuses on dynamics over a long-time period that significantly exceeds the initial training window. This transition is enabled by the adaptive transfer learnable mechanism, which combines the pre-trained neural foundation with a learnable, time-parameterized scaling function.
 This dual-stage approach preserves the symplectic geometry of the flow by leveraging prior physical knowledge to guide the adaptation of long-term trajectories.


\section{Numerical Results \label{sec:results}}

In this section, we present numerical tests to evaluate the performance of the proposed ATLAS-NN, with emphasis on transfer learning for different time intervals: we first train on a \emph{source} task defined on a \textbf{short-time interval} and then tune on a \emph{target} task defined on a \textbf{long-time interval}. Furthermore, we compare the performance of the proposed ATLAS-NN against several different models. To make the notations consistent, the acronyms for all models and numerical methods used throughout the paper are summarized in Table~\ref{tab:model-names}.

\begin{table}[H]
\centering
\caption{Acronyms for models and methods used in this paper.}
\label{tab:model-names}
\begin{tabular}{ll}
\hline
\textbf{Name} & \textbf{Description} \\
\hline
SE & Symplectic Euler integrator \\
HNN & Hamiltonian Neural Network proposed in reference\cite{mattheakis2022hamiltonian}; \eqref{eq:param_sol_ref} \\
ATLAS-NN & Time-Adaptive Transfer-Learnable Hamiltonian Neural Network \\
ATLAS-NN (tanh) & ATLAS-NN with temporal scaling function \eqref{eq:ft_tanh} \\
ATLAS-NN (exp) & ATLAS-NN with temporal scaling function \eqref{eq:ft_frac2}  \\
\hline
\end{tabular}
\end{table}

 \subsection{Nonlinear Oscillator}
As the first numerical example, we consider a one-dimensional nonlinear (anharmonic) oscillator \cite{mattheakis2022hamiltonian,bender1969anharmonic} to examine the effectiveness of the proposed models on a simple yet representative nonlinear Hamiltonian system. Despite its low-dimensional structure, this system retains essential nonlinear dynamical features, making it an appropriate benchmark for evaluating the ability of different methods to capture nonlinear effects and preserve the Hamiltonian structure without unnecessary complexity. Unlike the linear oscillator, the nonlinear oscillator exhibits a state-dependent frequency that changes with the oscillation amplitude, thus placing greater demands on both numerical integrators and neural network models. Meanwhile, since the dynamics remain non-chaotic, the system also provides a clean setting for systematically analyzing prediction errors and energy deviations.
The dynamics of the system are governed by the Hamiltonian, $H(x,p)$, defined as:
\begin{equation}
H(x,p) = \frac{p^{2}}{2} + \frac{x^{2}}{2} + \frac{x^{4}}{4},
\end{equation}
where $x$ denotes the generalized coordinate (position) and $p$ represents the conjugate momentum. In this formulation, both the mass and the natural frequency are set to unity. Since the Hamiltonian represents the total energy of the system, the associated equations of motion are derived from Hamilton's canonical equations:
\begin{equation}
\dot{x} = \frac{\partial H}{\partial p} = p, \qquad \dot{p} = -\frac{\partial H}{\partial x} = -(x + x^{3}),
\end{equation}
where the overdot denotes the derivative with respect to time $t$. We prescribe the initial state of the system at $t=0$ as:
\begin{equation}
x(0) = 1.3, \qquad p(0) = 1.
\end{equation}

\subsubsection{Short-time prediction\label{sec:no-short}}
To assess the fundamental accuracy and convergence properties of the proposed models, we first evaluate the system over a short time interval $[0,4\pi]$. 
This interval spans several oscillation periods and is used to learn the short-term dynamics of the nonlinear Hamiltonian system.

\paragraph{Numerical setting.}
For all experiments, $N=200$ collocation points are used for training. The regularized loss function in \eqref{eq:hnn_loss_ref} is adopted with regularization coefficient $\lambda_{reg}=1$. Three models, HNN, ATLAS-NN (tanh) and ATLAS-NN (exp) as summarized in Table \ref{tab:model-names}, are considered, each using the same fully connected neural network architecture with $2$ hidden layers and $50$ neurons per hidden layer. To ensure a consistent comparison, all models are trained for $10^5$ epochs with a learning rate of $8\times 10^{-3}$. In addition, to enforce the positivity constraints $\alpha>0$, $\beta>0$, and $m>0$ throughout training, these parameters are reparameterized via the Softplus function as
\begin{equation}
m=\mathrm{softplus}(m_{\mathrm{raw}}), \qquad
\alpha=\mathrm{softplus}(\alpha_{\mathrm{raw}}), \qquad
\beta=\mathrm{softplus}(\beta_{\mathrm{raw}}),
\end{equation}
where
\begin{equation}
\mathrm{softplus}(x)=\log(1+e^x).
\end{equation}
This reparameterization guarantees that the resulting parameters remain strictly positive during optimization.

\begin{figure}[H]
    \centering

    \begin{subfigure}[t]{0.6\textwidth}
        \centering
        \includegraphics[width=\linewidth]{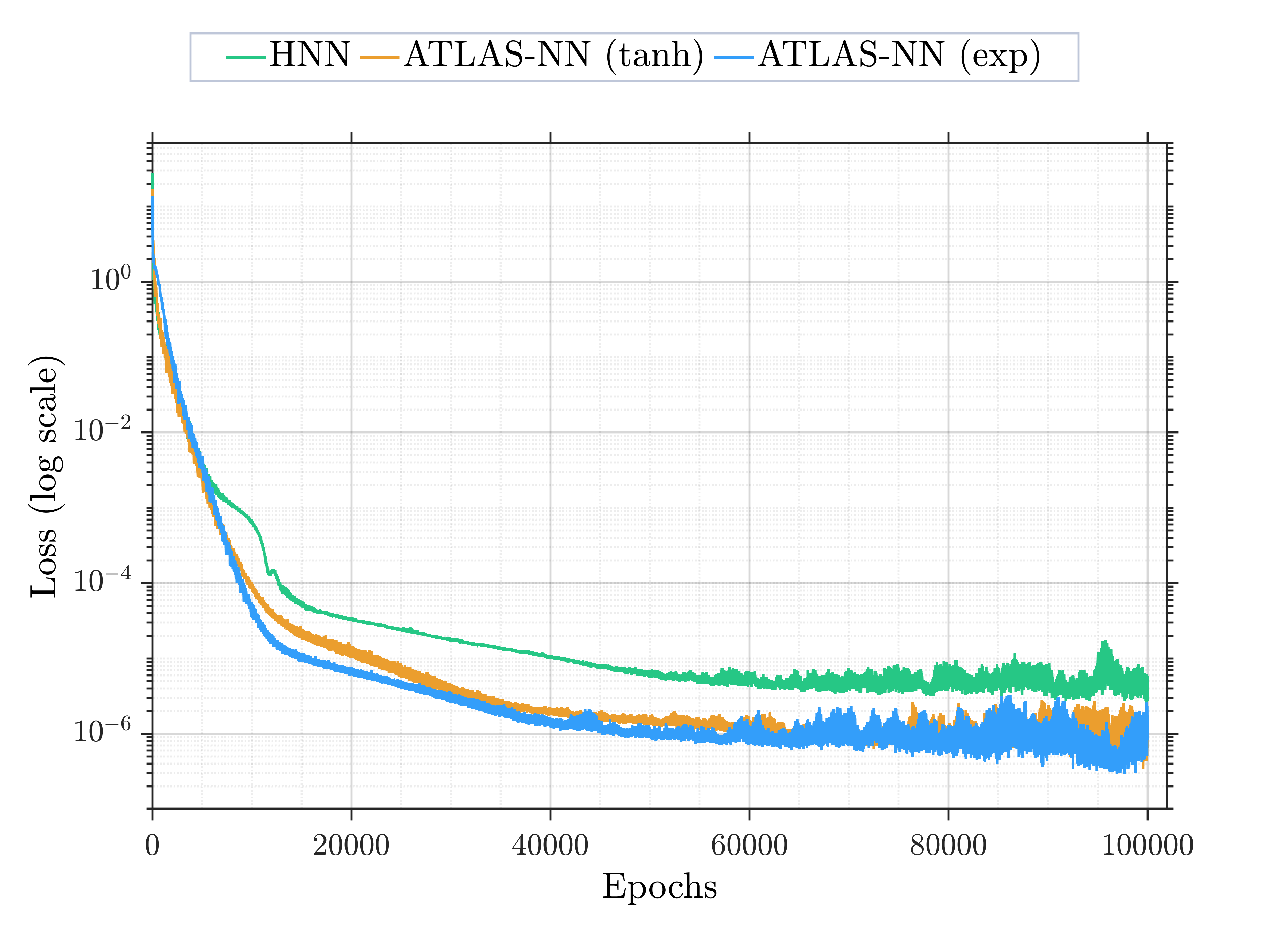}
        \caption{Training loss curves.}
        \label{fig:NL_loss}
    \end{subfigure}

    \vspace{0.5em}

    \begin{subfigure}[t]{0.49\textwidth}
        \centering
        \includegraphics[width=\linewidth]{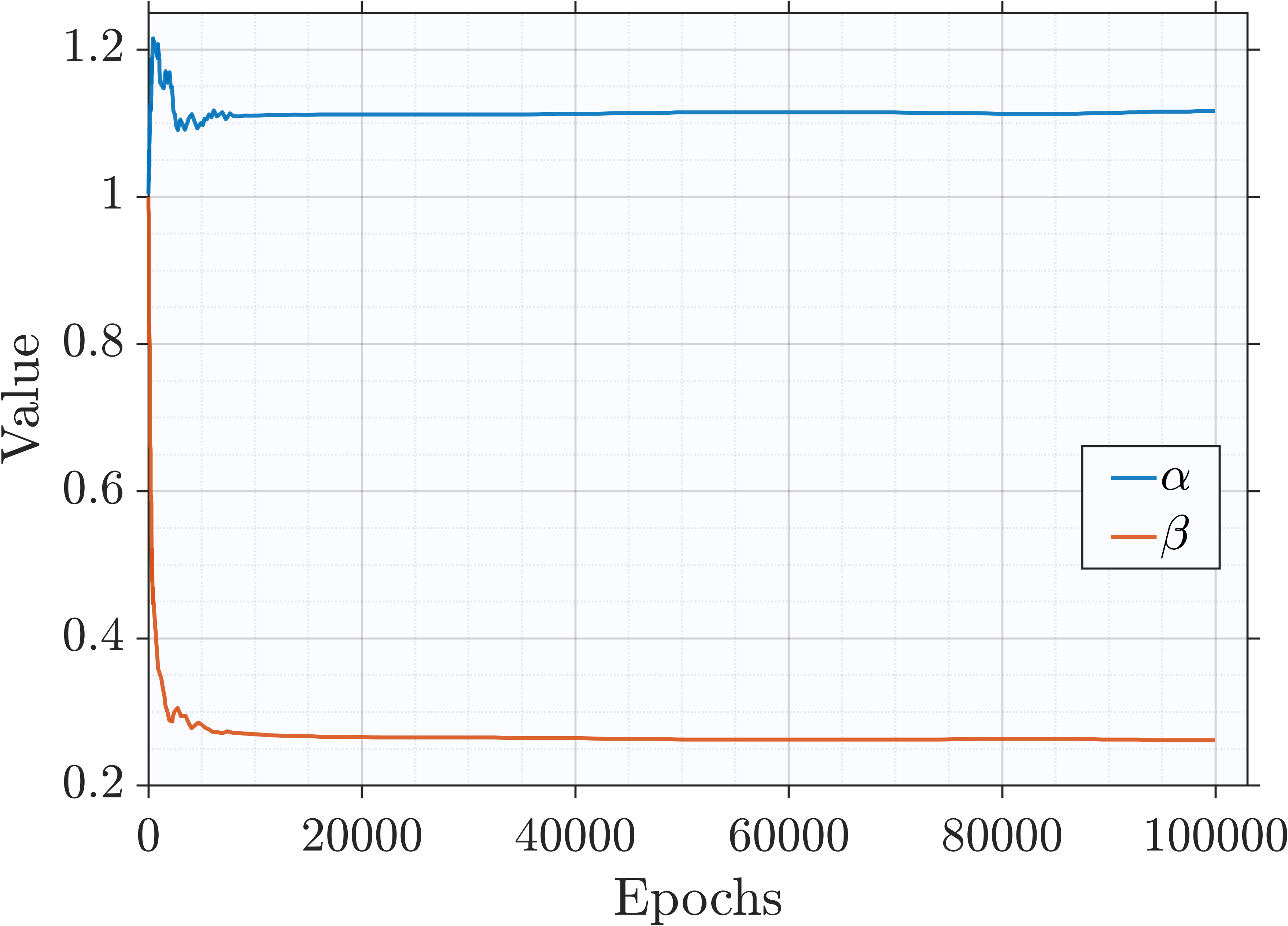}
        \caption{Evolution of $\alpha$ and $\beta$.}
        \label{fig:NL_alpha_beta}
    \end{subfigure}
    \hfill
    \begin{subfigure}[t]{0.49\textwidth}
        \centering
        \includegraphics[width=\linewidth]{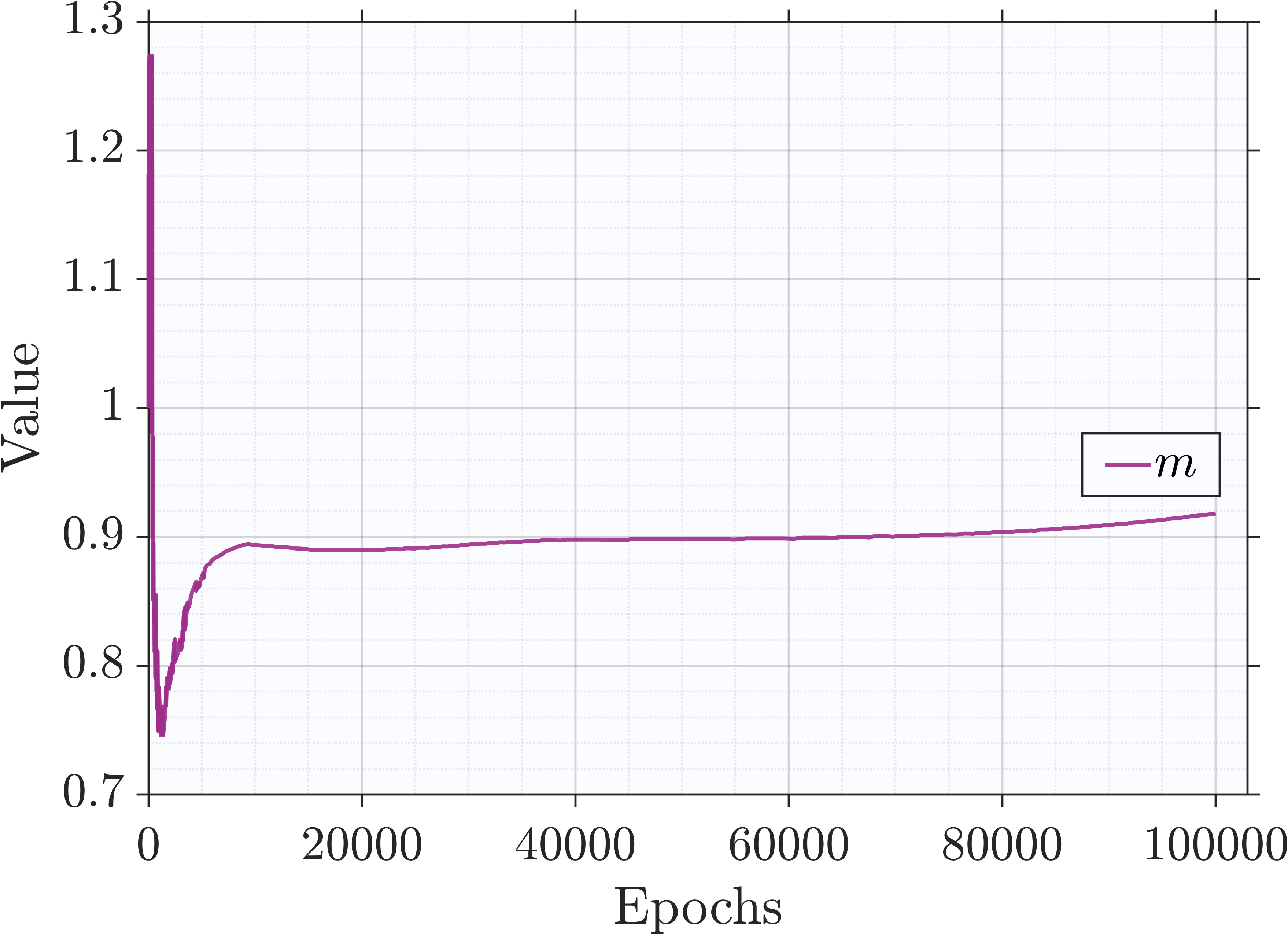}
        \caption{Evolution of $m$.}
        \label{fig:NL_m}
    \end{subfigure}

    \caption{Training behavior of the proposed models for the short-term nonlinear oscillator with short time time domain. 
    The top panel shows the training loss curves, where the time-adaptive ATLAS-NN models converge faster and reach lower final loss values than the standard HNN. 
    The bottom panels display the convergence of the learnable parameters $\alpha$, $\beta$, and $m$ during training.}
    \label{fig:NL_training}
\end{figure}

\paragraph{Training loss and parameter evolution.}
Figure~\ref{fig:NL_training} illustrates the training behavior of the HNN and the proposed ATLAS-NN models in the short time prediction. The loss curves show that the ATLAS-NNs converge more rapidly and reach lower final loss values than the standard HNN, indicating that the adaptive temporal weighting improves optimization efficiency during training. In addition, despite being initialized at $1.0$, these parameters gradually stabilize as training proceeds which suggests the convergence of the learnable temporal weighting function.

\paragraph{Quantitative comparison metrics.}
The predicted trajectories of all models listed in Table~\ref{tab:model-names} are compared against a high-accuracy reference solution, and the corresponding $L^2$ errors between the model predictions and the high-fidelity solution are computed as
\begin{equation}
\mathcal{E}_{L^2}
=
\left(
\int_0^T
\left\|
\mathbf{z}_{\mathrm{pred}}(t)-\mathbf{z}_{\mathrm{ref}}(t)
\right\|_2^2
\,dt
\right)^{1/2},
\end{equation}
where $\mathbf{z}(t)=[x(t),p(t)]^\top$ denotes the system state, $\mathbf{z}_{\mathrm{pred}}(t)$ is the predicted trajectory, and $\mathbf{z}_{\mathrm{ref}}(t)$ is the corresponding high-fidelity reference solution.
In addition, we use HNN as the baseline model and introduce the following metric to quantify the relative improvement of each ATLAS-NN:
\begin{equation}
\mathrm{Improvement}(\%)
=
\frac{|\mathcal{E}_{\mathrm{baseline}}-\mathcal{E}_{\mathrm{model}}|}
{\mathcal{E}_{\mathrm{baseline}}}\times 100\%,
\label{eq:improvement}
\end{equation}
where $\mathcal{E}_{\mathrm{baseline}}$ and $\mathcal{E}_{\mathrm{model}}$ represent the prediction errors of the baseline HNN and the corresponding model under consideration, respectively.

\begin{table}[H]
\centering
\renewcommand{\arraystretch}{1.25}
\setlength{\tabcolsep}{6pt}
\begin{tabular}{l l l
>{\centering\arraybackslash}p{2.4cm}
>{\centering\arraybackslash}p{1.8cm}
>{\centering\arraybackslash}p{2.6cm}
>{\centering\arraybackslash}p{1.8cm}}
\toprule
{Error} & 
{SE} &
{HNN} &
{ATLAS-NN (tanh)} & {Improvement} &
{ATLAS-NN (exp)} & {Improvement} \\
\midrule
$f(t)$
& N/A
& $1-e^{-t}$
& $\tanh(mt)$
& --
& $\displaystyle\frac{1-e^{-\alpha t}}{1+\beta e^{-\alpha t}}$
& -- \\
\midrule

$L^2(x)$
& $1.43\times10^{0}$
& $1.17\times10^{-2}$
& $1.12\times10^{-2}$
& $4.51\%$
& $4.31\times10^{-3}$
& \highlightgreen{$63.32\%$} \\

$L^2(p_x)$
& $3.57\times10^{0}$
& $2.79\times10^{-2}$
& $1.46\times10^{-2}$
& $47.66\%$
& $1.10\times10^{-2}$
& \highlightgreen{$60.60\%$} \\

MSE$(x)$
& $1.02\times10^{-2}$
& $6.90\times10^{-8}$
& $6.29\times10^{-8}$
& $8.82\%$
& $9.28\times10^{-9}$
& \highlightgreen{$86.56\%$} \\

MSE$(p_x)$
& $6.36\times10^{-2}$
& $3.90\times10^{-7}$
& $1.07\times10^{-7}$
& \highlightgreen{$72.65\%$}
& $6.04\times10^{-8}$
& \highlightgreen{$84.48\%$} \\
\bottomrule
\end{tabular}

\caption{
Quantitative comparison of Symplectic Euler (SE), Hamiltonian Neural Network (HNN), and time-adaptive HNN models, including ATLAS-NN (tanh) and ATLAS-NN (exp), for the source task.
Cells shaded in light green indicate relative improvements greater than $50\%$.
For the ATLAS-NN models, the time-adaptive functions are $f(t)=\tanh(mt)$ with $m=0.91799$, and
$f(t)=\frac{1-e^{-\alpha t}}{1+\beta e^{-\alpha t}}$ with $\alpha=1.11510$ and $\beta=0.26074$.
}
\label{tab:NL_errors}
\end{table}

\begin{figure}[H]
    \centering

    \begin{subfigure}[t]{0.8\linewidth}
        \centering
        \includegraphics[width=\linewidth,height=.55\linewidth]{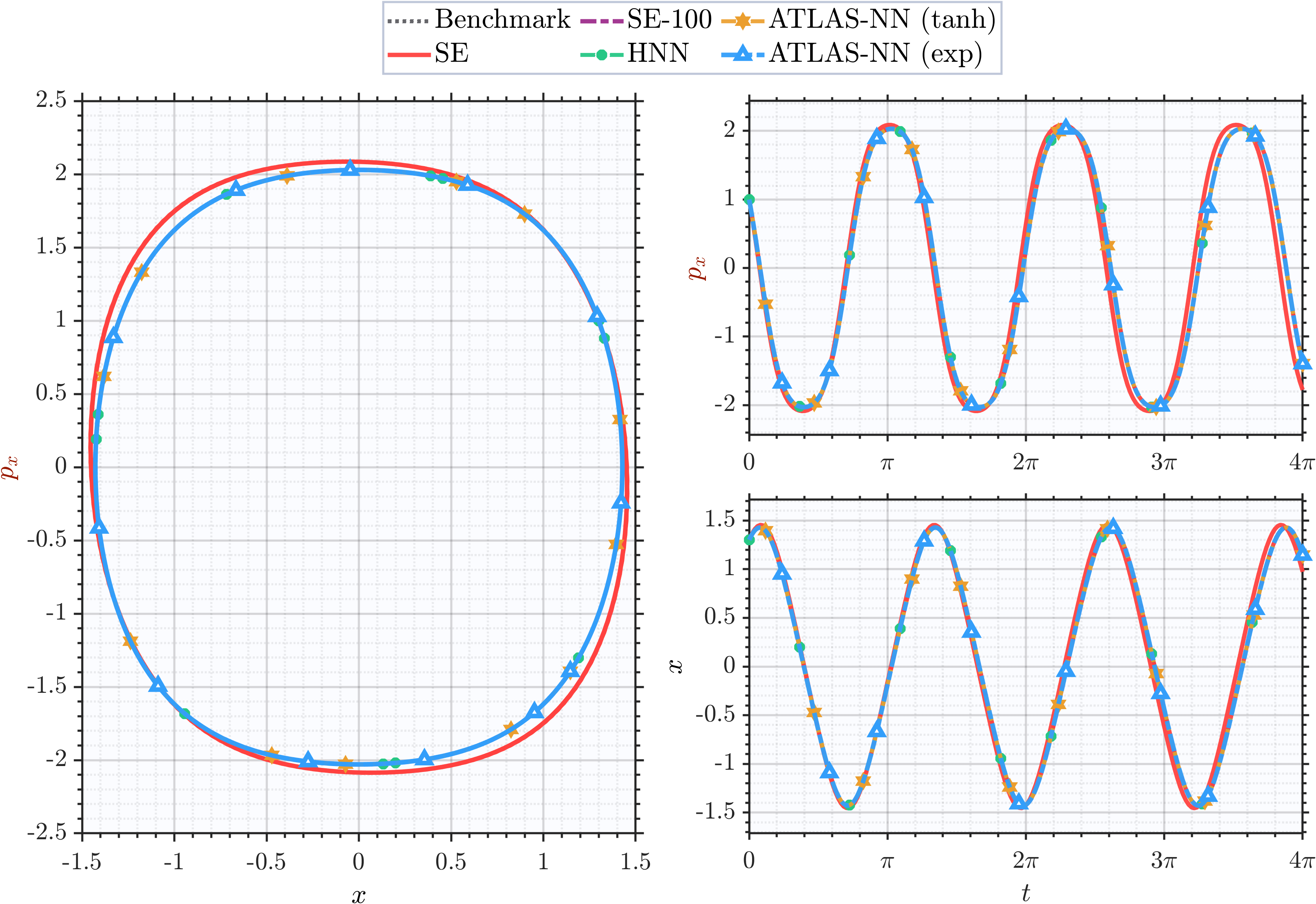}
        \caption{Trajectories in phase space and time evolution of the state variables.}
        \label{fig:NL_TrajectoriesCompare}
    \end{subfigure}

    \vspace{0.6em}

    \begin{subfigure}[t]{0.8\linewidth}
        \centering
        \includegraphics[width=\linewidth,height=.55\linewidth]{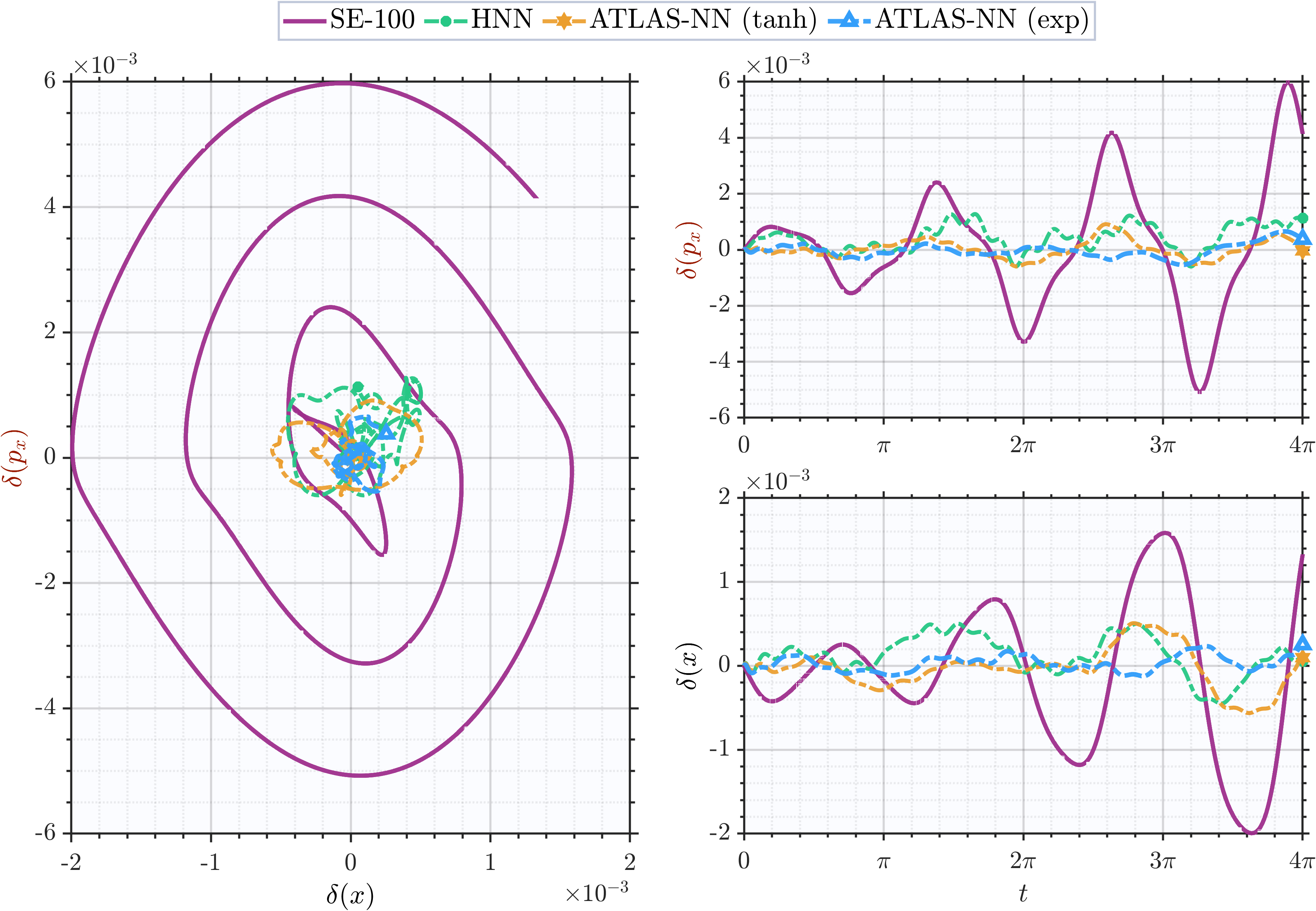}
        \caption{Comparison of prediction errors.}
        \label{fig:NL_error}
    \end{subfigure}

    \vspace{0.6em}

    \begin{subfigure}[t]{0.8\linewidth}
        \centering
        \includegraphics[width=\linewidth,height=.28\linewidth]{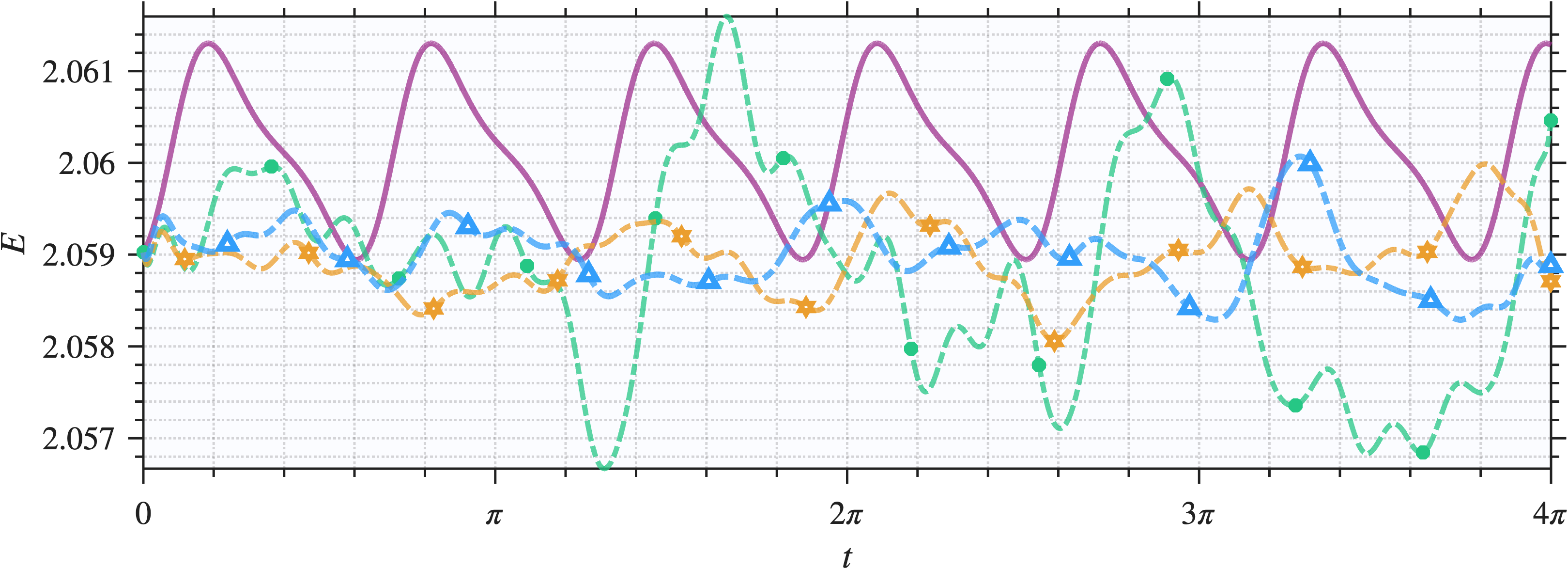}
        \caption{Energy evolution.}
        \label{fig:NL_energy}
    \end{subfigure}

    \caption{Comparison of model performance for the nonlinear oscillator over the source interval $[0,4\pi]$. The top panel shows the phase-plane trajectories together with the time evolution of the displacement $x$ and momentum $p$. The middle panel presents the corresponding prediction errors, and the bottom panel shows the energy evolution. SE-100 denotes the Symplectic Euler method using 100 times more training points than the neural-network-based models.}
    \label{fig:NL_results}
\end{figure}

\paragraph{Discussion.}
Table~\ref{tab:NL_errors} presents a quantitative comparison of Symplectic Euler (SE), Hamiltonian Neural Network (HNN), and time-adaptive HNN models, including ATLAS-NN (tanh) and ATLAS-NN (exp), for the short time prediction. As expected, the Symplectic Euler method yields the largest errors among all approaches, with $L^2$ errors of order $10^{0}$ for both $x$ and $p$, indicating limited accuracy under the present setting. In contrast, the HNN already provides a substantial improvement, reducing the errors by several orders of magnitude.  This observation is consistent with the results reported in \cite{mattheakis2022hamiltonian}.  On the other hand, both ATLAS-NN (tanh) and ATLAS-NN (exp) further improve the accuracy. In particular, ATLAS-NN (tanh) achieves modest improvement for $L^2(x)$, reducing the error from $1.17\times10^{-2}$ to $1.12\times10^{-2}$, while producing much more noticeable reduction for $L^2(p)$ and especially for $\mathrm{MSE}(p)$, where the relative improvement reaches $72.65\%$. 
Among all models, ATLAS-NN (exp) is the best for all reported metrics. For instance, compared with HNN, the $L^2$ error for $x$ is reduced by nearly one order of magnitude, corresponding to a $63.32\%$ improvement, while the $L^2$ error for $p$ decreases by $60.60\%$. Similar behavior is observed for the MSE metrics. In particular, the error in $x$ is reduced by nearly one order of magnitude, while the error in $p$ is also substantially decreased, by more than half relative to the HNN.
These results show that adaptive time-weighting improves the prediction accuracy of HNN, with the two-parameter exponential form being more effective than the one-parameter hyperbolic tangent form for this problem.
Figure~\ref{fig:NL_results} further confirms the quantitative findings in Table~\ref{tab:NL_errors}. In the top panel, all neural-network-based models closely follow the benchmark trajectory in both the phase plane and the time evolution of $x$ and $p$, whereas the Symplectic Euler method shows visibly larger deviations. The middle panel further confirms this behavior through the prediction error plots: SE exhibits the largest errors over time, while HNN significantly reduces these deviations. Both two different ATLAS-NNs provide additional improvement, with ATLAS-NN (exp) showing the smallest overall error magnitude and the closest agreement with the benchmark throughout the entire interval. The bottom panel shows the corresponding energy evolution, where the adaptive models maintain the Hamiltonian more accurately than the standard HNN, again with ATLAS-NN (exp) displaying the most stable and accurate behavior.

\subsubsection{Transfer learning: from source task to target task \label{sec:no-transfer}}

In this section, we investigate the effectiveness of transfer learning in enhancing the long-term predictive performance of the Nonlinear Oscillator. Specifically, we examine whether the proposed \textbf{ATLAS-NN} framework offers superior adaptability and efficiency when transitioning from short-term source tasks to long-time integration. 
In particular, the source task is chosen to be consistent with the short-term evaluation in Section \ref{sec:no-short} over the interval $[0,4\pi]$, while the target task is conducted over the extended long-time interval $[0, 20\pi]$.

\paragraph{Numerical setting.}
To accommodate the increased complexity of the long-term dynamics in the target task, the architecture is expanded to $80$ neurons per hidden layer. Under the transfer learning framework, this target model inherits the foundational features from the source task by loading the pre-trained weights and the optimized time-adaptive parameters, thereby leveraging the previously learned inductive bias for the extended interval.
Optimization is performed using the Adam optimizer with a learning rate of $5\times10^{-3}$, and training terminates once the total loss reaches a prescribed tolerance of $4\times10^{-5}$.

\paragraph{Training strategies.}
To evaluate the impact of adaptive transfer learnable mechanism, we compare two different :
\begin{itemize}
\item \textbf{Direct Training (Baseline):} To establish a performance benchmark, the standard HNN is trained from scratch directly on the long-time interval $[0, 20\pi]$. This baseline model uses $N=500$ uniformly sampled collocation points and random initialization which serves a baseline model that without the transfer learning scheme. We denote the basline model as baseline HNN. Hereafter, this reference model is denoted as the \textbf{baseline HNN}.
    \item \textbf{Transfer Learning (from Source Task to Target Task):} To evaluate the generalizability of the learned dynamics, all models (HNN, ATLAS-NN (tanh), and ATLAS-NN (exp)) are trained using a transfer learning process:
    \begin{itemize}
        \item \textit{Source Task:} The model is trained on the source interval $[0, 4\pi]$, as detailed in the previous section, until the training loss reaches the prescribed tolerance of $4\times10^{-5}$.
        \item \textit{Target Task:} The optimized neural network weights and time-adaptive parameters ($\alpha, \beta, m$) from Stage 1 are loaded as the initialization. The model is then fine-tuned on the long-time target interval $[0, 20\pi]$ using $N=500$ points, maintaining the same stopping criterion. 
\item \textit{Nomenclature:} The models utilizing this transfer learning approach are hereafter denoted as \textbf{transfer HNN}, \textbf{transfer ATLAS-NN (tanh)}, and \textbf{transfer ATLAS-NN (exp)}.        
    \end{itemize}
\end{itemize}
\paragraph{Transfer learning in ATLAS-NN.}
A defining feature of the \textit{\textbf{ATLAS-NN}} framework is the treatment of the learnable time-adaptive parameters ($\alpha, \beta, m$) during the transition between tasks. In the target task, the earnable time-adaptive parameters optimized during the source task are frozen, ensuring that only the neural network weights are fine-tuned for the long-time regime. By fixing these parameters, we isolate the influence of the transferred temporal bias, allowing for a direct assessment of how the learned inductive bias contributes to transfer efficiency and long-term structural stability.

\begin{figure}[H]
    \centering
    \includegraphics[width=0.7\linewidth]{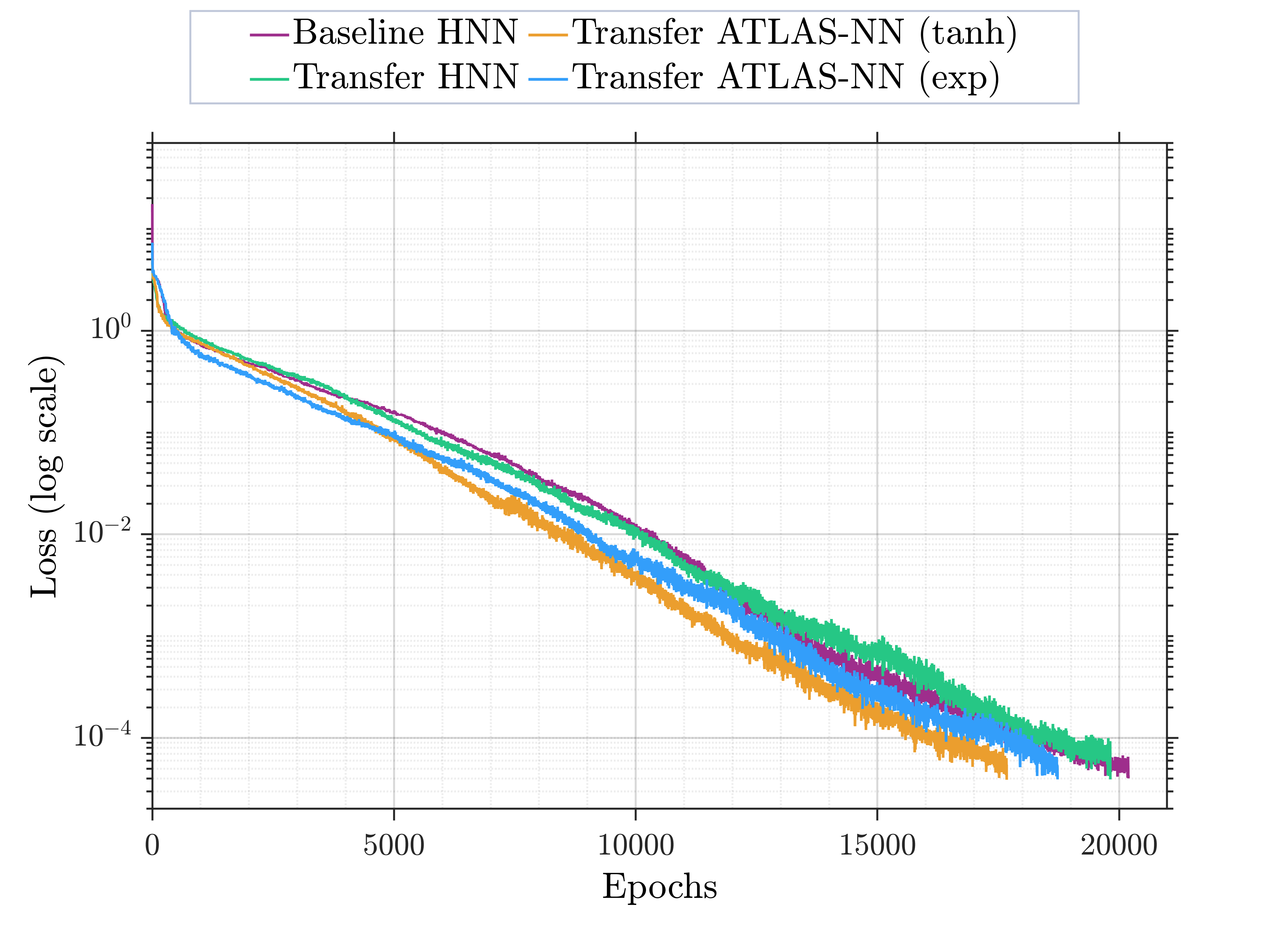}
    \caption{
Training loss comparison of different models, baseline HNN, transfer HNN, transfer ATLAS-NN(tanh) and transfer ATLAS-NN(exp,) for the nonlinear oscillator over $[0, 20\pi]$. 
A reference loss threshold of $4\times 10^{-5}$ is indicated to evaluate convergence efficiency. 
Both ATLAS-NN variants leveraging the adaptive transfer learnable mechanism demonstrate significantly faster convergence than the standard HNN. 
}
\label{fig:target_loss_comparison}
\end{figure}
\paragraph{Results and discussion.}
The convergence behavior illustrated in Figure~\ref{fig:target_loss_comparison} compares the training efficiency of the various models during the target task over the long-time interval $[0, 20\pi]$. While the \textbf{transfer HNN} (green) exhibits a convergence profile similar to the \textit{baseline HNN} (purple),
both ATLAS-NN variants demonstrate a clear advantage. Specifically, the \textit{transfer ATLAS-NN (exp)} (blue) maintains a consistently lower loss throughout the training process, while the \textit{transfer ATLAS-NN (tanh)} (orange) achieves the prescribed $4 \times 10^{-5}$ threshold most efficiently in approximately $17,000$ epochs. These results indicate that the time-adaptive parameters ($\alpha, \beta, m$) optimized during the source task provide a robust inductive bias for the nonlinear oscillator. By freezing these parameters during the target task, the \textit{ATLAS-NN} framework stabilizes the gradient flow and achieves the target accuracy with reduced computational training cost compared to the standard HNN.

The numerical results summarized in Table \ref{tab:no-error-transfer} provide a quantitative assessment of the proposed framework's performance over the long-time interval. A primary observation is the stark contrast between the neural network approaches and the traditional \textit{Symplectic Euler (SE)} method. While SE is a structure-preserving integrator, it nonetheless exhibits significant error accumulation over the extended $[0, 20\pi]$ horizon, with an $L^2(x)$ error of $2.72 \times 10^{1}$. In comparison, even the \textit{baseline HNN} reduces this error by approximately two orders of magnitude, underscoring the inherent advantage of continuous neural representations in capturing Hamiltonian vector fields.
Within the neural network architectures, the impact of the transfer learning framework is evident. The \textit{transfer HNN} yields a moderate improvement over the baseline, suggesting that weight initialization from a source task provides a better starting point for the optimizer. However, the most significant gains are observed in the \textit{transfer ATLAS-NN} models. Specifically, the \textit{transfer ATLAS-NN (exp)} achieves the highest precision, reducing the $L^2(x)$ error by nearly an order of magnitude and the MSE by nearly two orders of magnitude relative to the baseline HNN. 
This hierarchical improvement, from SE to HNN, and finally to ATLAS-NN, highlights the efficacy of the adaptive transfer learnable mechanism. While standard HNNs rely solely on fixed coordinates, the ATLAS-NN variants leverage the time-adaptive parameters ($\alpha, \beta, m$) learned in the source task. By freezing these parameters during the target task, the model maintains a superior "temporal bias" that prevents the phase drift and energy inaccuracies typically associated with long-time integration. Consequently, the ATLAS-NN models not only outperform traditional numerical methods by several orders of magnitude but also offer a more robust and computationally efficient alternative to standard Hamiltonian neural architectures.

\begin{figure}[H]
    \centering

    \begin{subfigure}[t]{.9\linewidth}
        \centering
    \includegraphics[width=1\linewidth]{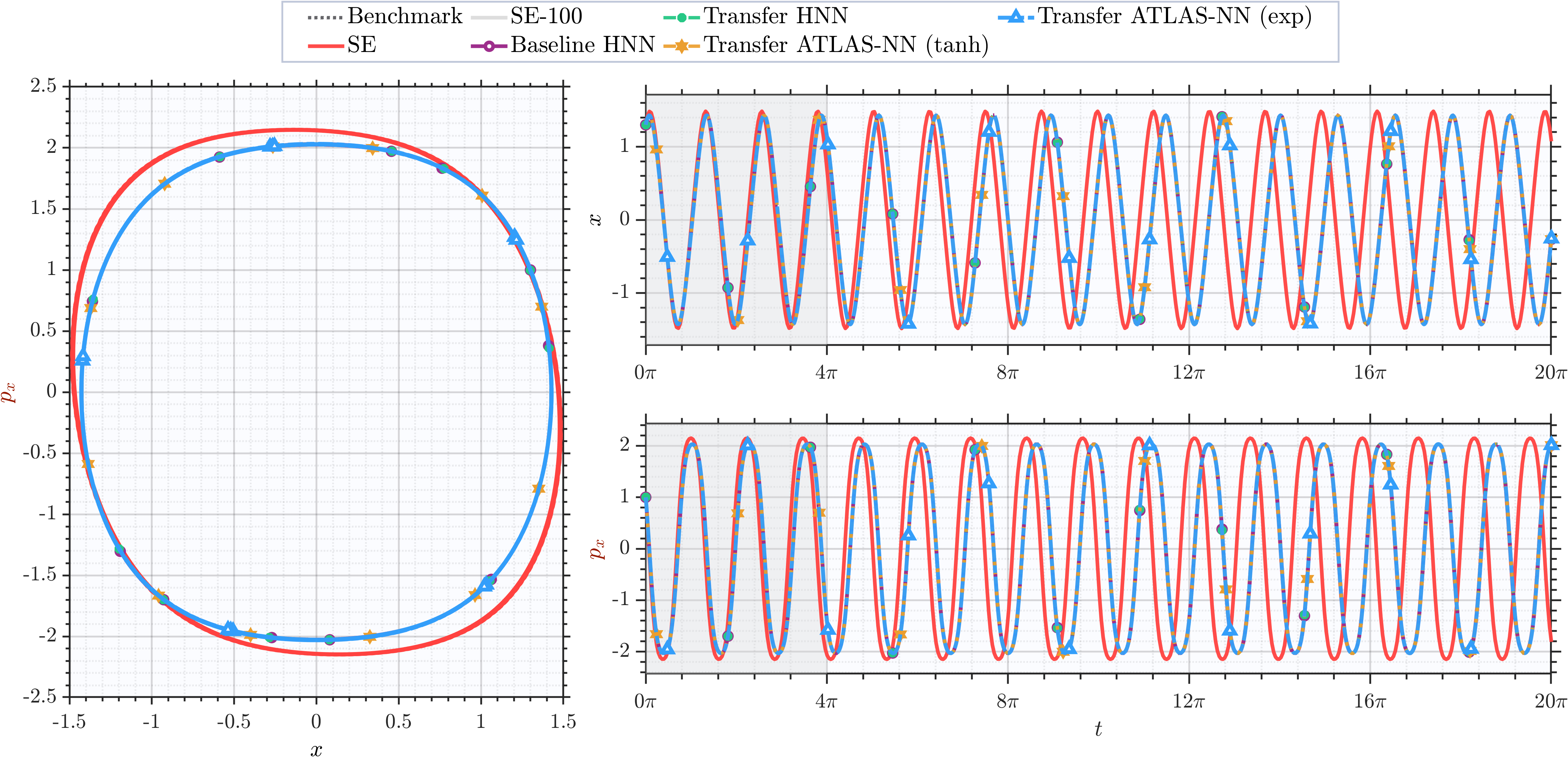}
        \caption{Trajectories in phase space and time evolution of the state variables.}
        \label{fig:NL_TrajectoriesCompare_transfer}
    \end{subfigure}

    \vspace{0.6em}

    \begin{subfigure}[t]{.9\linewidth}
        \centering
        \includegraphics[width=\linewidth]{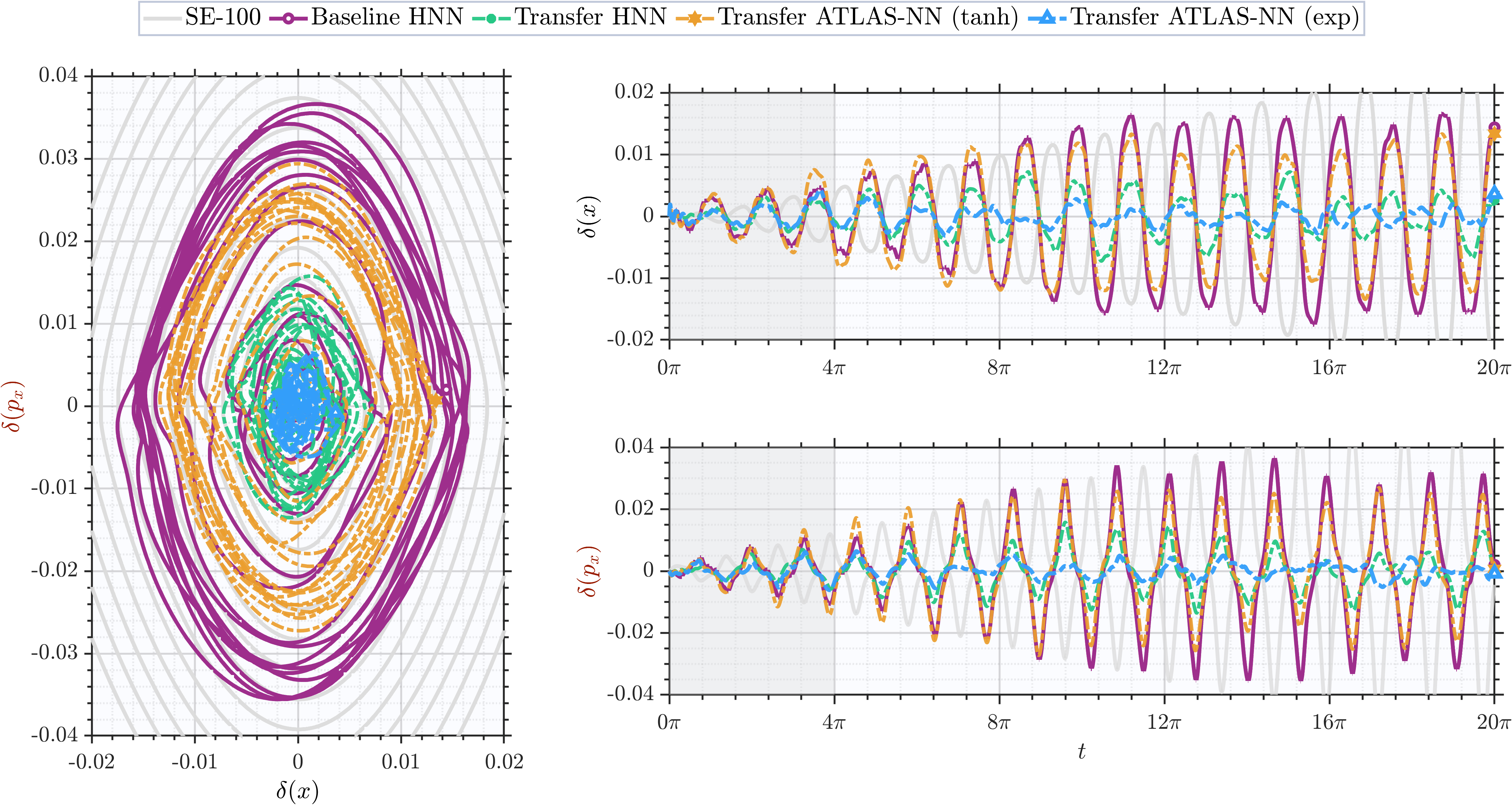}
        \caption{Comparison of prediction errors.}
        \label{fig:NL_error_transfer}
    \end{subfigure}

    \vspace{0.6em}

    \begin{subfigure}[t]{.9\linewidth}
        \centering
        \includegraphics[width=\linewidth]{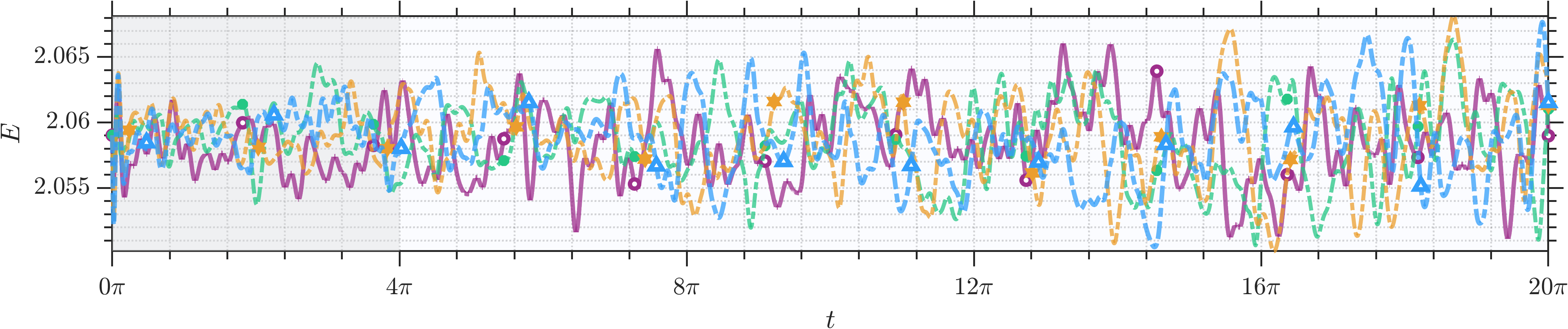}
        \caption{Energy evolution.}
        \label{fig:NL_energy_transfer}
    \end{subfigure}

\caption{Comparison of long-time model performance for the nonlinear oscillator over the target interval $[0, 20\pi]$, where the shaded region $[0, 4\pi]$ denotes the source interval. Panel (a) illustrates the phase-space trajectories alongside the temporal evolution of state variables $x$ and $p_x$. Panel (b) presents the corresponding prediction errors $\delta(x)$ and $\delta(p_x)$. Panel (c) displays the energy evolution, highlighting the structure-preserving capabilities of the different methods. Here, SE-100 denotes the Symplectic Euler method utilizing 100 times the collocation points of the neural network models.}    \label{fig:NL_results_transfer}
\end{figure}

Following the quantitative analysis in Table~\ref{tab:no-error-transfer}, Figure~\ref{fig:NL_results_transfer} provides a qualitative assessment of the long-term structural stability across the full interval $[0, 20\pi]$. While all models accurately track the trajectory within the shaded source interval $[0, 4\pi]$, the \textit{baseline HNN} and \textit{SE-100} exhibit visible phase drift and expanding error envelopes (Panel b) as time progresses. In contrast, the \textit{ATLAS-NN} variants maintain significantly higher fidelity, with prediction errors remaining nearly an order of magnitude lower than the baseline by the end of the simulation. 
The energy evolution in Panel (c) further shows the effectiveness of the adaptive transfer mechanism. While the traditional Symplectic Euler method and standard HNN show fluctuating energy profiles in the long-time regime, the \textit{transfer ATLAS-NN} models preserve the Hamiltonian within the right variation. This suggests that the fixed time-adaptive parameters ($\alpha, \beta, m$) act as a corrective inductive bias, effectively suppressing the accumulation of numerical errors and ensuring the trajectory remains on the correct energy manifold over extended temporal horizons.
\begin{table}[H]
\centering
\renewcommand{\arraystretch}{1.25}
\setlength{\tabcolsep}{6pt}
\begin{tabular}{
>{\centering\arraybackslash}p{1cm} 
>{\centering\arraybackslash}p{1.2cm} 
>{\centering\arraybackslash}p{1.8cm}
>{\centering\arraybackslash}p{1.8cm}
>{\centering\arraybackslash}p{1.2cm} 
>{\centering\arraybackslash}p{1.8cm} 
>{\centering\arraybackslash}p{1.2cm} 
>{\centering\arraybackslash}p{1.8cm} 
>{\centering\arraybackslash}p{1.2cm} 
}
\toprule 
\textbf{Error} &\textbf{SE}& \textbf{HNN\qquad (Baseline)} &\textbf{HNN \qquad (Transfer)}&\textbf{Imp.} & \textbf{ATLAS-NN \ (Transfer)} &\textbf{Imp.} & \textbf{ATLAS-NN \ (Transfer)} &\textbf{Imp.}\\
\midrule
$f(t)$
&N/A
& $1-e^{-t}$
& $1-e^{-t}$& --
& $\tanh(mt)$& --
& $\displaystyle\frac{1-e^{-\alpha t}}{1+\beta e^{-\alpha t}}$& --
 \\
\midrule 
$L^2(x)$   &2.72$\times10^{1}$ & 6.68 $\times10^{-1}$   & 2.24 $\times10^{-1}$ & $66.45\%$& 5.55$\times10^{-1}$ & $16.90\%$ & 9.73$\times10^{-2}$ & \highlightgreen{$85.45\%$} \\
$L^2(p)$   &4.53$\times10^{1}$ & 1.08   &  3.96$\times10^{-1}$ & $63.33\%$& 8.81$\times10^{-1}$ & $18.39\%$ & 1.69$\times10^{-1}$ & \highlightgreen{$84.38\%$} \\
MSE$(x)$  & 1.48 & 8.93$\times10^{-5}$ &1.01$\times10^{-5}$ & \highlightgreen{$88.75\%$}& 6.17$\times10^{-5}$ & $30.95\%$& 1.89$\times10^{-6}$ & \highlightgreen{$97.88\%$} \\
MSE$(p)$   &4.12 & 2.33$\times10^{-4}$ &3.14$\times10^{-5}$& \highlightgreen{$86.55\%$} & 1.55$\times10^{-4}$ & $33.39\%$& 5.69$\times10^{-6}$ & \highlightgreen{$97.56\%$} \\
\midrule
\textbf{Epochs } &N/A& 20191 &19818 & $1.85\%$& 17673& $12.47\%$ & 18727 & $7.25\%$ \\
\bottomrule 
\end{tabular}
\caption{
Error comparison between the ATLAS-NN transfer learning models. Percentage improvements indicate relative error reduction with respect to the HNN baseline.
Cells shaded in light celadon green indicate improvements exceeding $80\%$.
For the ATLAS-NN models, the time-adaptive functions are $f(t)=\tanh(mt)$ with $m=0.87440$, and
$f(t)=\frac{1-e^{-\alpha t}}{1+\beta e^{-\alpha t}}$ with $\alpha=1.08015$ and $\beta=0.22404$.
\label{tab:no-error-transfer}
}
\end{table}

\subsection{H\'enon--Heiles System \label{sec:hh}}
As a second numerical benchmark, we consider the H\'enon--Heiles (HH) Hamiltonian system. Originally proposed to model stellar motion within an axisymmetric galactic potential, this system has become a foundational model in nonlinear dynamics and chaos theory. Despite its relatively simple polynomial form, the HH system exhibits a complex transition from quasi-periodic motion to formal chaos as the total energy increases. This characteristic makes it an ideal test case for evaluating the long-term stability and structural preservation of the proposed \textit v{ATLAS-NN} framework in higher-dimensional phase spaces.
The Hamiltonian of the H\'enon--Heiles system, representing the total energy of a particle in an axisymmetric potential, is defined as:
\begin{equation}
H(x, y, p_x, p_y) = \underbrace{\frac{1}{2}(p_x^2 + p_y^2)}_{\text{Kinetic Energy}} + \underbrace{\frac{1}{2}(x^2 + y^2) + \left(x^2 y - \frac{1}{3}y^3\right)}_{V(x, y) \text{ (Potential Energy)}},
\end{equation}
where $(x, y)$ denote the Cartesian coordinates (positions) and $(p_x, p_y)$ are their corresponding conjugate momenta. In this formulation, the cubic terms $x^2 y$ and $-\frac{1}{3}y^3$ introduce the essential nonlinearity that drives the transition toward chaotic dynamics. The resulting Hamilton's canonical equations of motion, which govern the temporal evolution of the system, are given by:
\begin{equation}
\begin{aligned}
\dot{x} &= \frac{\partial H}{\partial p_x} = p_x, & \dot{y} &= \frac{\partial H}{\partial p_y} = p_y, \\
\dot{p}_x &= -\frac{\partial H}{\partial x} = -x - 2xy, & \dot{p}_y &= -\frac{\partial H}{\partial y} = -y - x^2 + y^2.
\end{aligned}
\end{equation}

We prescribe the initial condition at 
$t=0$ by
\begin{equation}
x_0(0)=0.3,\qquad 
y_0(0)=-0.3,\qquad 
p_x(0)=0.3,\qquad 
p_y(0)=0.15.
\end{equation}

\subsubsection{Short-time Prediction \label{sec:hh-short}}



Consistent with Section \ref{sec:no-short}, we first evaluate the HH system over a short time interval $[0,6\pi]$, which corresponds to approximately $0.069\, t_{\max}$ Lyapunov times.  

\paragraph{Numerical setting.}
For all experiments, $N=500$ training points are used for training. The regularized loss function in \eqref{eq:hnn_loss_ref} is adopted with regularization 
coefficient $\lambda=0.5$. Three models, HNN, ATLAS-NN (tanh) and ATLAS-NN (exp) as summarized in Table \ref{tab:model-names}, are considered using the same fully connected neural network architecture with $2$ hidden layers and $80$ neurons  per hidden layer. To ensure a consistent comparison, all models are trained 
for $3 \times 10^{4}$ epochs with a learning rate of $3 \times 10^{-3}$.
In addition, to enforce the positivity constraints $\alpha>0$, $\beta>0$, and $m>0$ throughout training, we use the same strategy as in Section \ref{sec:no-short}. 

\paragraph{Training loss and parameter evolution.}
Figure~\ref{fig:NL_training} illustrates the training behavior of the HNN and the proposed ATLAS-NN models in the short time prediction. Consistent with the nonlinear oscillator test case, the loss curves shows that the ATLAS-NNs converge more rapidly and reach lower terminal loss values than the standard HNN. Furthermore, the lower panels track the evolution of 
the learned parameters $\alpha$, $\beta$, and $m$, all initialized at $1.0$. These parameters gradually stabilize as training proceeds which suggests the convergence of the learnable temporal weighting function.

\paragraph{Quantitative comparison metrics.}
The predicted trajectories produced by all models listed in Table \ref{tab:HH_errors} are compared with a high-accuracy reference solution. We employ the same quantitative error metrics and baseline model as described in Section \ref{sec:no-short}.



\begin{table}[H]
\centering
\renewcommand{\arraystretch}{1.25}
\setlength{\tabcolsep}{6pt}
\begin{tabular}{l l l
>{\centering\arraybackslash}p{2.4cm}
>{\centering\arraybackslash}p{1.8cm}
>{\centering\arraybackslash}p{2.6cm}
>{\centering\arraybackslash}p{1.8cm}}
\toprule
{Error} & 
{SE} &
{HNN} &
{ATLAS-NN} & {Improvement} &
{ATLAS-NN} & {Improvement} \\
\midrule
$f(t)$
& N/A
& $1-e^{-t}$
& $\tanh(mt)$
& --
& $\displaystyle\frac{1-e^{-\alpha t}}{1+\beta e^{-\alpha t}}$
& -- \\
\midrule
$L^2(x)$
& $2.40\times10^{-1}$
& $1.84\times10^{-3}$
& $1.26\times10^{-3}$
& $31.59\%$
& $5.60\times10^{-4}$
& \highlightgreen{$69.48\%$} \\

$L^2(y)$
& $1.62\times10^{-1}$
& $2.71\times10^{-3}$
& $1.19\times10^{-3}$
& \highlightgreen{$56.19\%$}
& $7.63\times10^{-4}$
& \highlightgreen{$71.90\%$} \\

$L^2(p_x)$
& $1.56\times10^{-1}$
& $1.84\times10^{-3}$
& $1.09\times10^{-3}$
& $40.73\%$
& $1.14\times10^{-3}$
& $38.17\%$ \\

$L^2(p_y)$
& $9.02\times10^{-2}$
& $1.45\times10^{-3}$
& $1.07\times10^{-3}$
& $25.98\%$
& $7.68\times10^{-4}$
& $47.06\%$ \\

MSE$(x)$
& $1.16\times10^{-4}$
& $6.74\times10^{-9}$
& $3.15\times10^{-9}$
& \highlightgreen{$53.24\%$}
& $6.27\times10^{-10}$
& \highlightgreen{$90.72\%$} \\

MSE$(y)$
& $5.24\times10^{-5}$
& $1.47\times10^{-8}$
& $2.83\times10^{-9}$
& \highlightgreen{$80.80\%$}
& $1.16\times10^{-9}$
& \highlightgreen{$92.10\%$} \\

MSE$(p_x)$
& $4.88\times10^{-5}$
& $6.76\times10^{-9}$
& $2.38\times10^{-9}$
& \highlightgreen{$64.85\%$}
& $2.59\times10^{-9}$
& \highlightgreen{$61.77\%$} \\

MSE$(p_y)$
& $1.63\times10^{-5}$
& $4.19\times10^{-9}$
& $2.29\times10^{-9}$
& $45.23\%$
& $1.18\times10^{-9}$
& \highlightgreen{$71.80\%$} \\
\bottomrule
\end{tabular}

\caption{
Quantitative comparison of Symplectic Euler (SE), Hamiltonian Neural Network (HNN), and time-adaptive HNN models, including ATLAS-NN (tanh) and ATLAS-NN (exp), for the short time prediction.
Cells shaded in light green indicate relative improvements greater than $50\%$.
For the ATLAS-NN models, the time-adaptive functions are $f(t)=\tanh(mt)$ with $m=0.3964$, and
$f(t)=\frac{1-e^{-\alpha t}}{1+\beta e^{-\alpha t}}$ with $\alpha=0.6049$ and $\beta=1.0076$.
}
\label{tab:HH_errors}
\end{table}

\begin{figure}[H]
    \centering

    \begin{subfigure}[t]{0.6\textwidth}
        \centering
        \includegraphics[width=\linewidth]{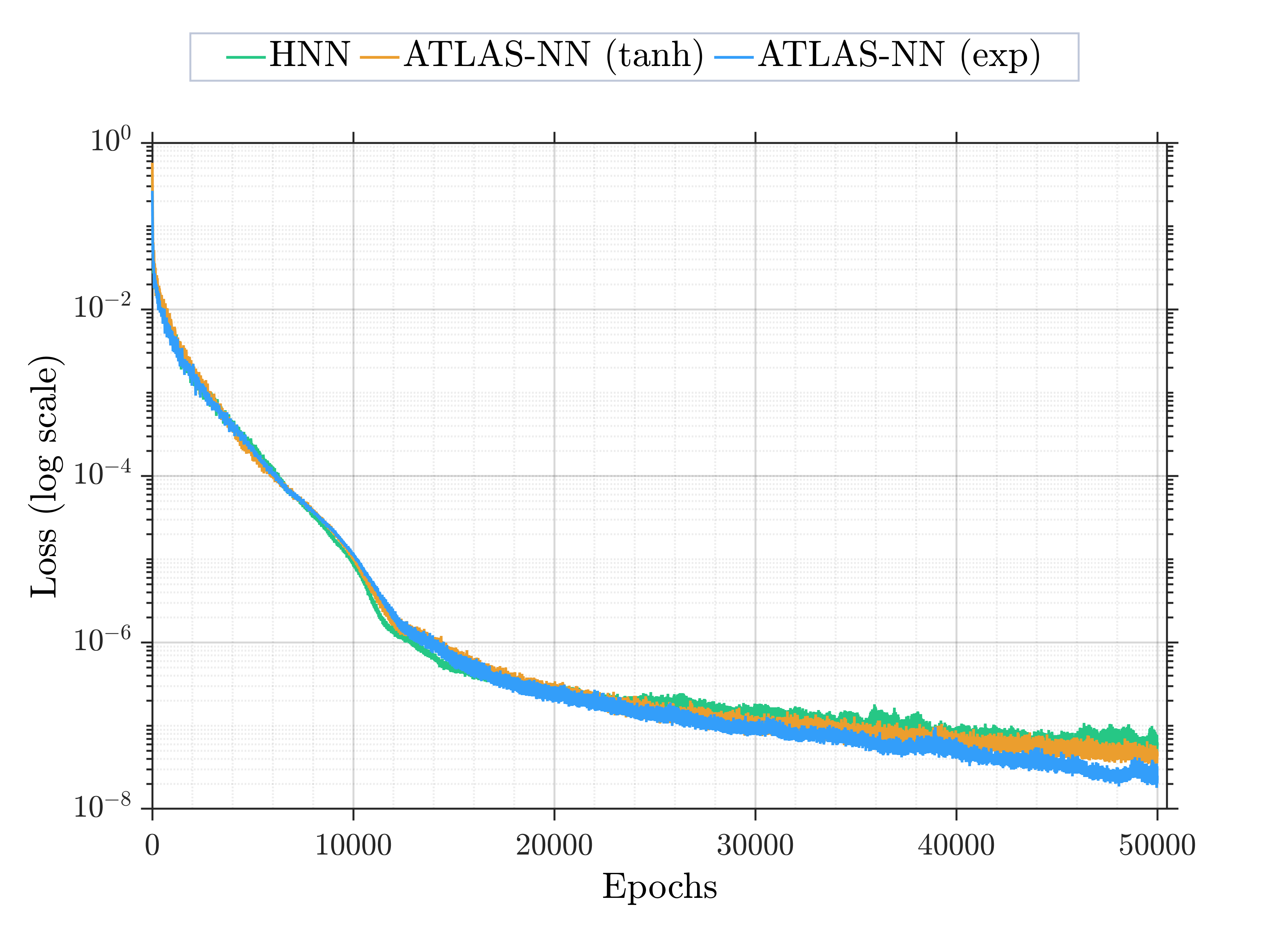}
        \caption{Training loss curves.}
        \label{fig:HH_loss}
    \end{subfigure}

    \vspace{0.5em}

    \begin{subfigure}[t]{0.49\textwidth}
        \centering
        \includegraphics[width=\linewidth]{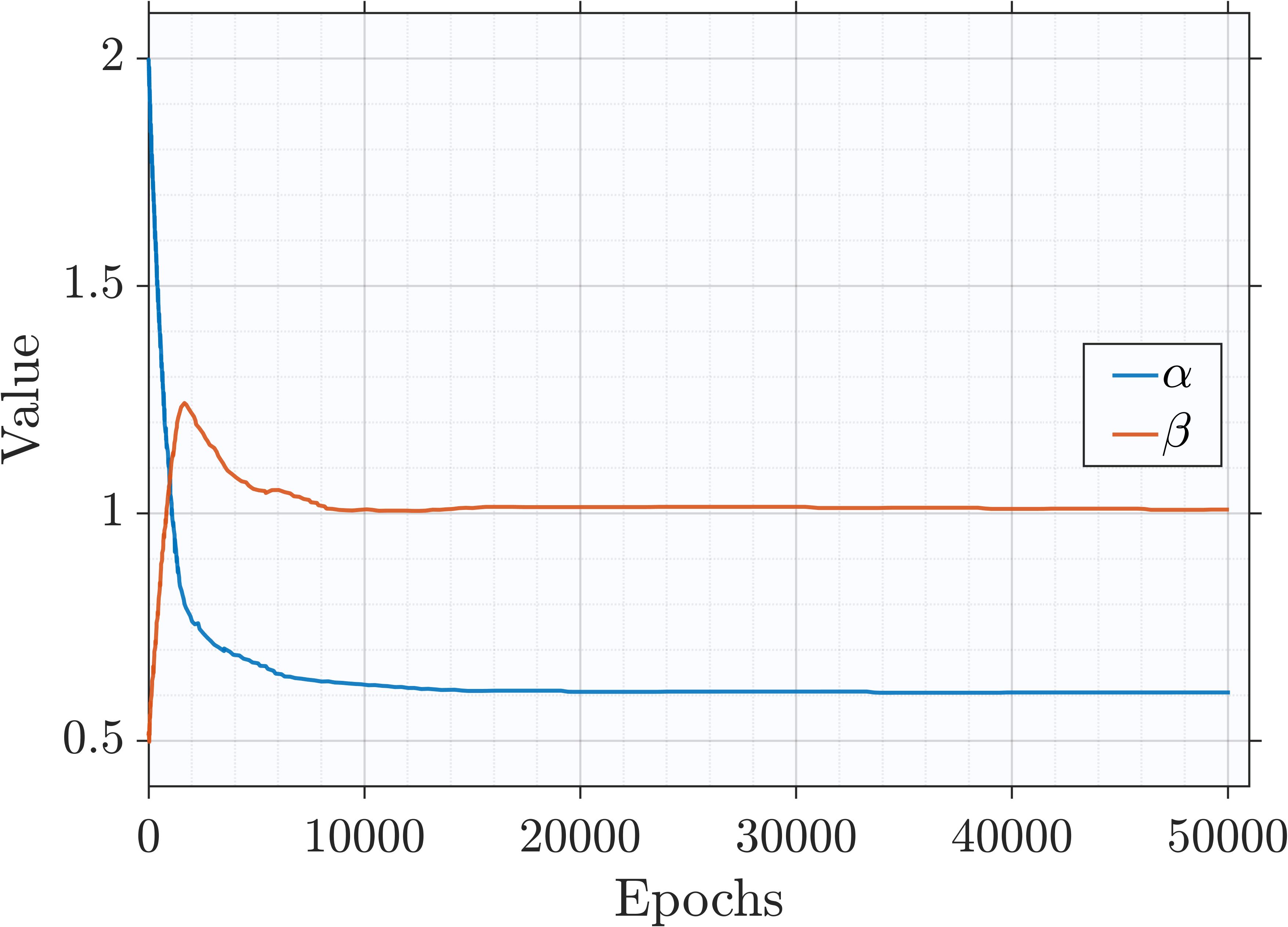} 
        \caption{Evolution of $\alpha$ and $\beta$.}
        \label{fig:HH_alpha_beta}
    \end{subfigure}
    \hfill
    \begin{subfigure}[t]{0.49\textwidth}
        \centering
        \includegraphics[width=\linewidth]{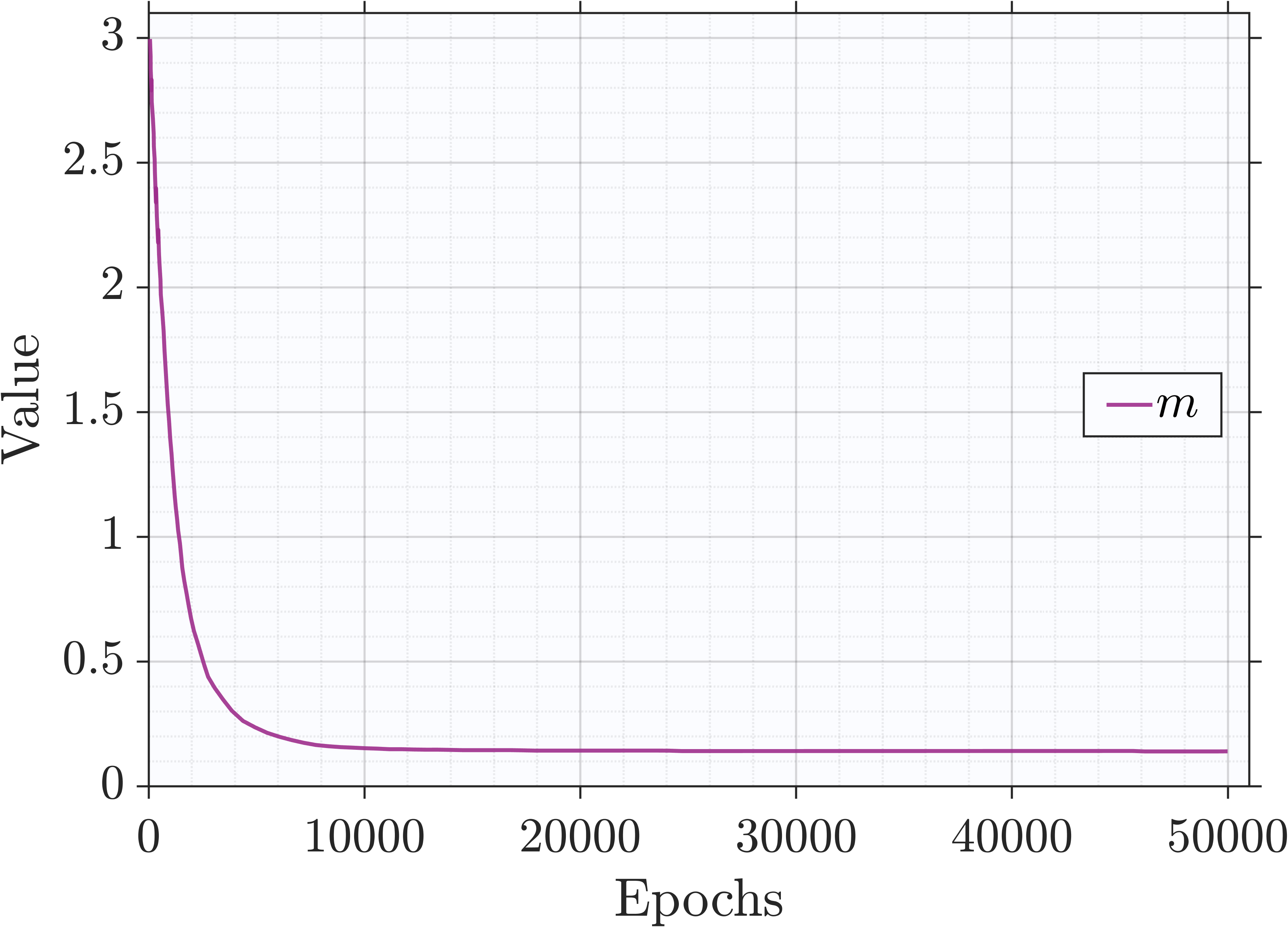}
        \caption{Evolution of $m$.}
        \label{fig:HH_m}
    \end{subfigure}

    \caption{Training behavior of the proposed models for the short-term the H\'enon--Heiles system short time time domain. 
    The top panel shows the training loss curves, where the time-adaptive ATLAS-NN models converge faster and reach lower final loss values than the standard HNN. 
    The bottom panels display the convergence of the learnable parameters $\alpha$, $\beta$, and $m$ during training.}
    \label{fig:HH_training}
\end{figure}

\begin{figure}[H]
    \centering
    \includegraphics[width=.9\linewidth]{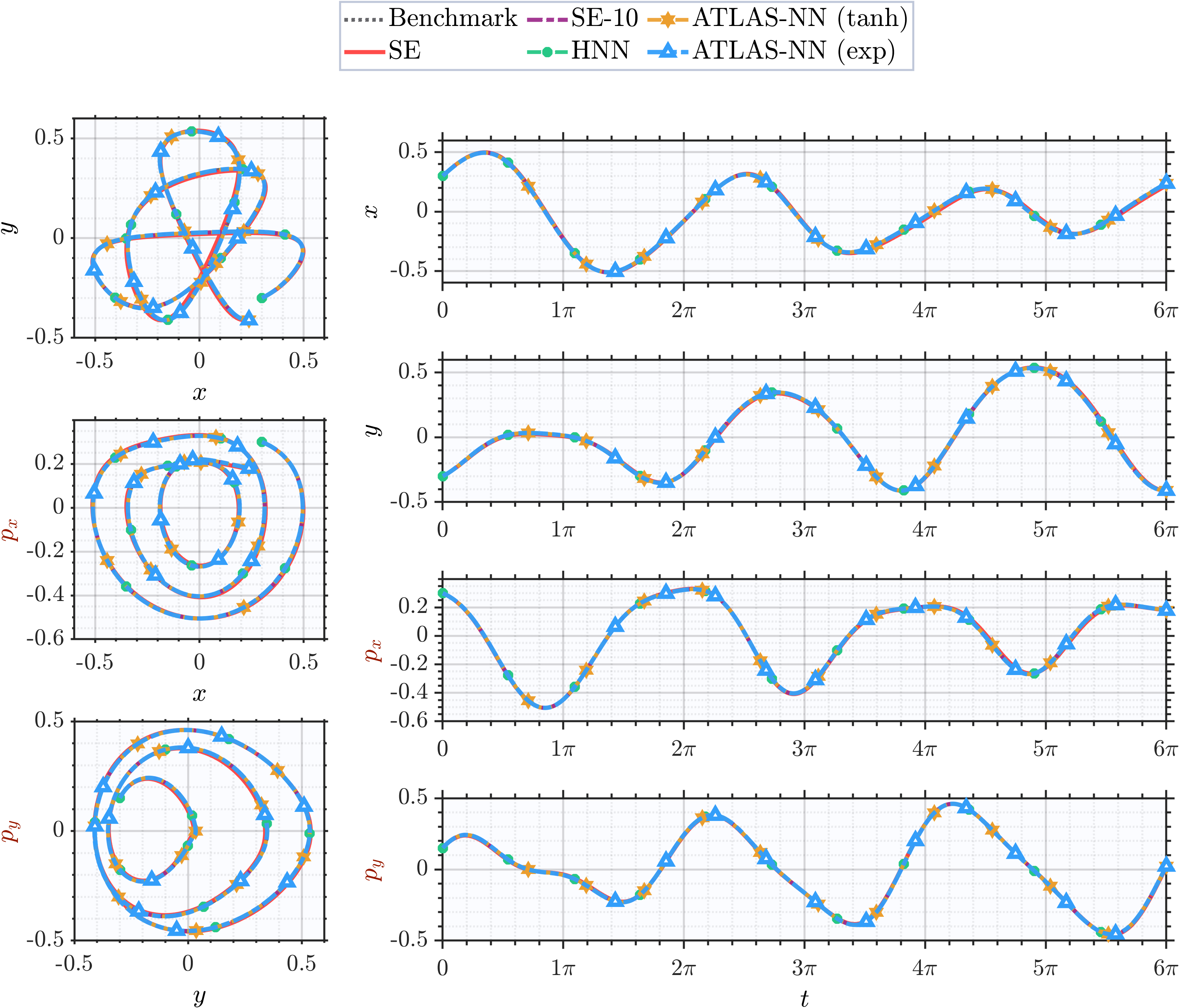}
    \caption{Trajectories of the Hénon--Heiles system for different models in the short time $[0,6\pi]$ prediction. 
The three left panels show the phase space trajectories in the $(x,y)$, $(x,p_x)$, and $(y,p_y)$ planes, and the four panels on the right illustrate the time evolution for each state variable: $x$, $y$, $p_x$, and $p_y$.}
    \label{fig:HH_traj}
\end{figure}
\begin{figure}[H]
    \centering

    \begin{subfigure}[t]{.9\linewidth}
        \centering
        \includegraphics[width=\linewidth]{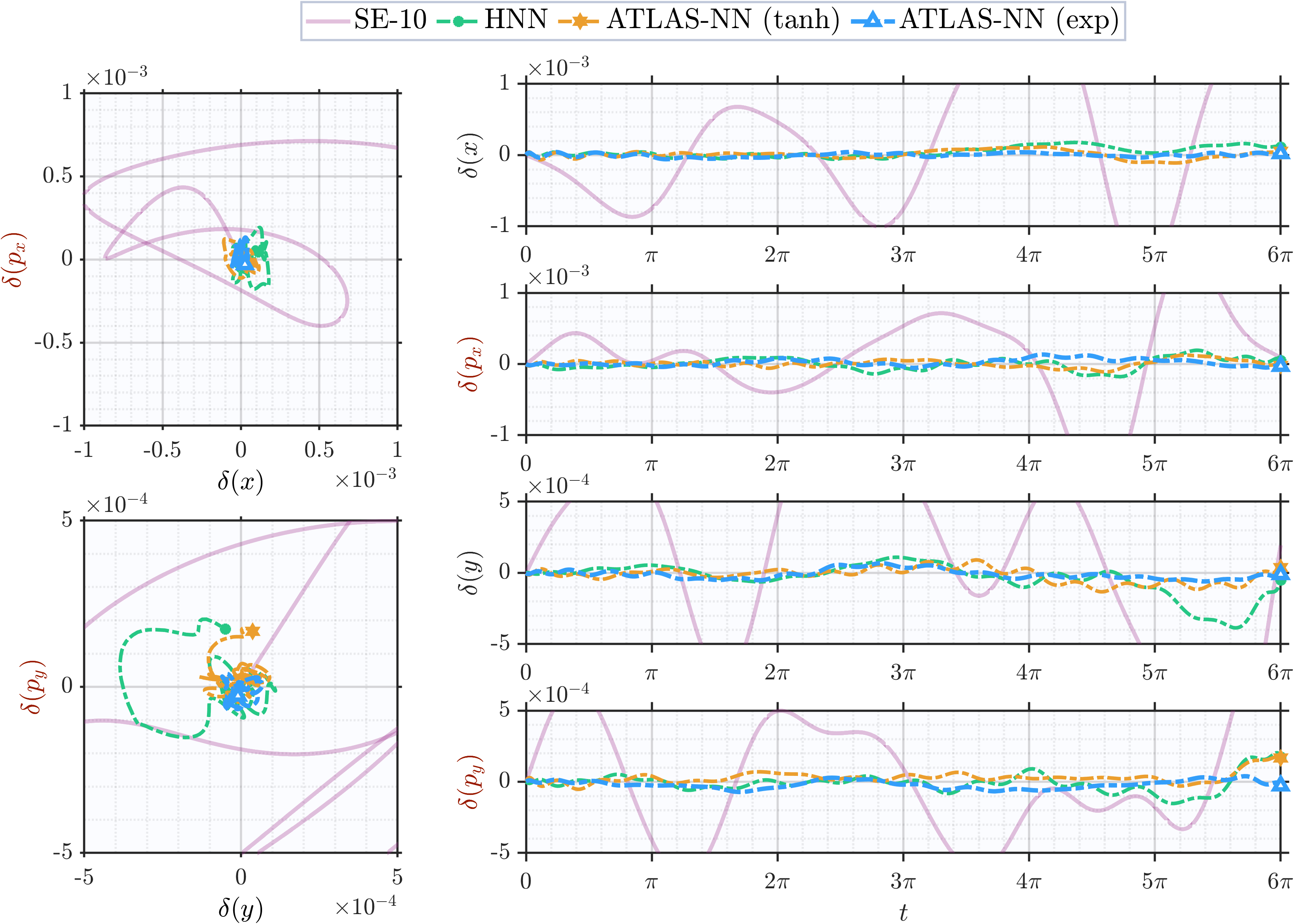}
        \caption{Error comparison}
        \label{fig:HH_error}
    \end{subfigure}

    \vspace{0.6em}

    \begin{subfigure}[t]{.9\linewidth}
        \centering
        \includegraphics[width=\linewidth]{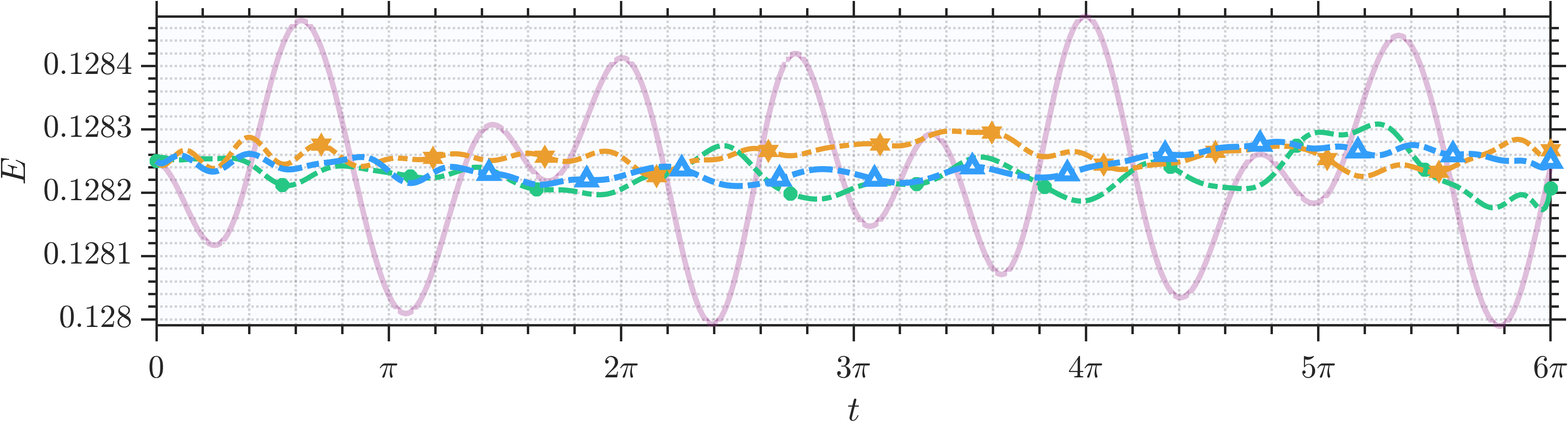}
        \caption{Energy evolution}
        \label{fig:HH_energy}
    \end{subfigure}

    \caption{
    Time evolution of error and energy of the Hénon--Heiles system for different models in the short time $[0,6\pi]$ prediction. The left subpanels of Panel (a) show the phase space trajectories in the $(x,y)$, $(x,p_x)$, and $(y,p_y)$ planes, and the four right subpanels of Panel (a) illustrate the time evolution for each state variable: $x$, $y$, $p_x$, and $p_y$; Panel (b) shows the time evolution of the energy for the H\'enon-Heiles system.}
    \label{fig:HH_ErrorCompare}
\end{figure}

\paragraph{Discussion.} Table~\ref{tab:HH_errors} presents a quantitative comparison of Symplectic Euler (SE), Hamiltonian Neural Network (HNN), and ATLAS-NN (tanh) and ATLAS-NN (exp) for the short time prediction.
As expected, the Symplectic Euler method yields the largest errors across all coordinates. In contrast, the HNN provides a substantial improvement which reduces these errors by \textit{several orders of magnitude}. This observation confirms the baseline accuracy of neural-network-based Hamiltonian solvers for this short time prediction and is consistent with the results reported in Sections \ref{sec:no-short}.
Both ATLAS-NN models provide further improvements over the standard HNN. In particular, ATLAS-NN (tanh) reduces the error by \textit{more than half} for several metrics, including $L^2(y)$ and the MSE of the position coordinates. However, among all models, \textit{ATLAS-NN (exp)} is the most accurate for nearly all quantitative metrics. For instance, compared with the standard HNN, the MSE for the position coordinates is reduced by \textit{nearly an order of magnitude}. Furthermore, the errors in the momentum coordinates are also substantially decreased, often by \textit{more than half} relative to the HNN. These results also highlighted by the shaded cells in Table~\ref{tab:HH_errors}, show that the ATLAS-NN(exp) is exceptionally effective, consistently outperforming the ATLAS-NN(tanh) for this problem. 
The accuracy of the models is further examined via the trajectories in Figure \ref{fig:HH_traj}. The phase space projections (left) and temporal evolutions (right) show that all models, including \textit{HNN}, \textit{SE}, and \textit{ATLAS-NN} models, capture the system's main characteristic orbital geometry. On the other hand, 
the temoporal evolution of error and energy of all models, illustrated in Figure \ref{fig:HH_ErrorCompare}, provide a deeper look into their robustness. As shown in Figure \ref{fig:HH_error}, the \textit{SE-10} model exhibits significant oscillatory error growth and phase drift in both coordinate and momentum deviations. In contrast, the \textit{HNN} and \textit{ATLAS-NN} variants maintain stable, low-magnitude errors throughout the integration period $t \in [0, 6\pi]$. The \textit{ATLAS-NN} models, in particular, demonstrate the most consistent accuracy in terms of error accumulation which keeps the deviations tightly bounded near the origin.
Energy conservation, a definitive metric for Hamiltonian systems, with its temporal evolution is shown in Figure \ref{fig:HH_energy}. While the \textit{SE-10} model fails to preserve the system energy $E$ that results in large-amplitude fluctuations around the benchmark value, the \textit{HNN} and \textit{ATLAS-NN} models exhibit remarkable energy stability. These models maintain the Hamiltonian nearly constant at $E \approx 0.1282$ with only minor, bounded oscillations. This indicates the intrinsic advantage of the \textit{ATLAS-NN} and \textit{HNN} frameworks in respecting physical constraints, which provides the long-time simulation possible. In addition, the SE model is omitted from the figure due to its significantly larger error magnitude, which would obscure the details of the other models.

\subsubsection{Transfer learning: from source task to target task}

\paragraph{Numerical setting.}
In this section, we evaluate the effectiveness of transfer learning in enhancing the long-term predictive performance of the H\'{e}non-Heiles system. We specifically examine whether the \textbf{ATLAS-NN} framework facilitates adaptability when transitioning from short-term source tasks to long time domain. Following the setup in Section \ref{sec:no-short}, the source task is defined over the interval $[0,6\pi]$, while the target task encompasses an extended long-time horizon of $[0, 24\pi]$.
To address the increased complexity of the target task's long-term dynamics, each hidden layer is expanded to $80$ neurons. The target model inherits foundational features and time-adaptive parameters from the pre-trained source task, leveraging the learned inductive bias for the extended interval. Optimization is conducted using the Adam optimizer with a learning rate of $5\times10^{-3}$, with training concluding upon reaching a loss tolerance of $4\times10^{-5}$.

\paragraph{Training strategies.}
Following the methodology used for the nonlinear oscillator in Section \ref{sec:no-transfer}, we evaluate the impact of the adaptive transfer mechanism by comparing two training strategies:
\begin{itemize}
    \item \textbf{Direct Training (Baseline):} A standard HNN is trained from scratch on the target interval $[0, 24\pi]$ using $N=1000$ uniformly sampled collocation points and random initialization. This reference model is denoted as the \textbf{baseline HNN}.
    
    \item \textbf{Transfer Learning:} The models (\textit{HNN}, \textit{ATLAS-NN (tanh)}, and \textit{ATLAS-NN (exp)}) undergo two standard transfer learning tasks:
    \begin{itemize}
        \item \textit{Source task:} Training on the interval $[0, 4\pi]$ until a loss of $4\times10^{-5}$ is reached.
        \item \textit{Target task:} The pre-trained weights and time-adaptive parameters ($\alpha, \beta, m$) are loaded as initializations for fine-tuning on the $[0, 24\pi]$ interval with $N=1000$ points.
    \end{itemize}
    The resulting models are denoted as \textbf{transfer HNN} and \textbf{transfer ATLAS-NN}.
\end{itemize}

\paragraph{Transfer Learning in ATLAS-NN}
During the target task, the time-adaptive parameters ($\alpha, \beta, m$) optimized in the source task remain frozen. This isolates the transferred temporal bias, allowing us to directly assess how the learned inductive bias contributes to transfer efficiency and structural stability in the long-time regime.

\begin{figure}[H]
    \centering
    \includegraphics[width=0.75\linewidth]{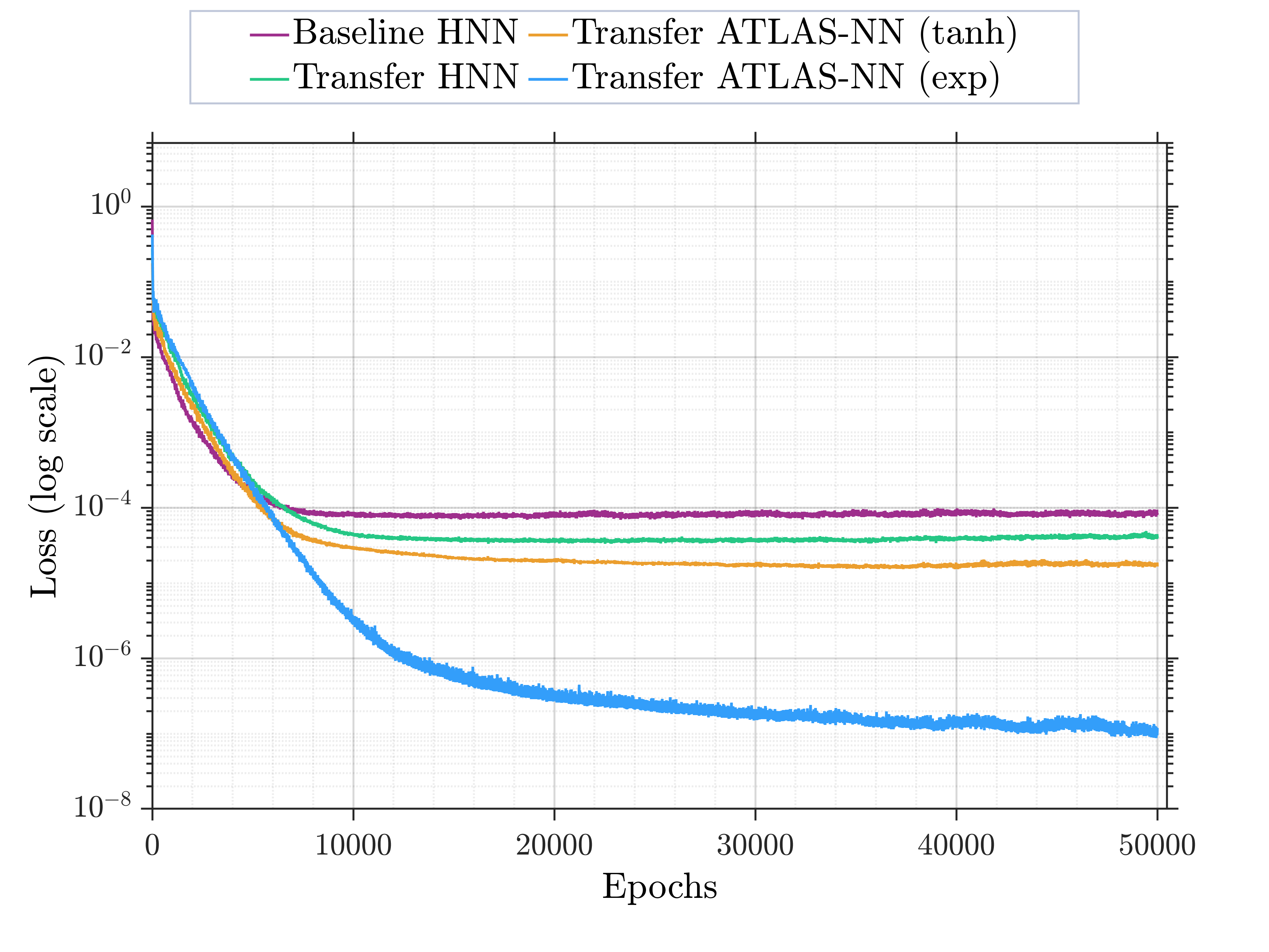}
    \caption{Training loss comparison of transfer learning models for the H\'enon--Heiles system. While transfer learning moderately improves HNN, the optimization quickly saturates. The ATLAS-NN with $f(t)=\tanh(mt)$ converges faster, whereas the ATLAS-NN with 
$f(t)=\frac{1-e^{-\alpha t}}{1+\beta e^{-\alpha t}}$ maintains a sustained decay of the training loss.
}
    \label{fig:HH_tran_loss}
\end{figure}

\paragraph{Results and discussions.}
The convergence behavior illustrated in Figure~\ref{fig:HH_tran_loss} compares the training efficiency of different models during the target task over the extended interval $[0, 24\pi]$. While the \textit{baseline HNN} (purple) plateaus at a loss of approximately $10^{-4}$, the transfer learning models exhibit significantly improved convergence. Although the \textit (green) shows a similar initial trajectory to the baseline, it achieves a lower final error, suggesting that pre-trained weight initialization provides a more favorable starting point in the optimization landscape.
The \textit{ATLAS-NN} models show a clear faster and more accurate convergence. Specifically, the \textit{transfer ATLAS-NN (exp)} (blue) exhibits the most robust performance, maintaining a consistently steeper descent and reaching a final loss nearly three orders of magnitude lower than the baseline. Meanwhile, the \textit{transfer ATLAS-NN (tanh)} (orange) achieves the prescribed $4 \times 10^{-5}$ threshold most efficiently, surpassing the target accuracy in approximately $10,000$ epochs.

\begin{table}[H]
\centering
\renewcommand{\arraystretch}{1.25}
\setlength{\tabcolsep}{6pt}
\begin{tabular}{
>{\centering\arraybackslash}p{1cm} 
>{\centering\arraybackslash}p{1.8cm} 
>{\centering\arraybackslash}p{1.8cm}
>{\centering\arraybackslash}p{1.8cm}
>{\centering\arraybackslash}p{1cm} 
>{\centering\arraybackslash}p{1.8cm} 
>{\centering\arraybackslash}p{1cm} 
>{\centering\arraybackslash}p{1.8cm} 
>{\centering\arraybackslash}p{1cm} 
}
\toprule 
\textbf{Error} &\textbf{SE}& \textbf{HNN\qquad (Baseline)} &\textbf{HNN \qquad (Transfer)}&\textbf{Imp.} & \textbf{ATLAS-NN \ (Transfer)} &\textbf{Imp.} & \textbf{ATLAS-NN \ (Transfer)} &\textbf{Imp.}\\
\midrule
$f(t)$
&N/A
& $1-e^{-t}$
& $1-e^{-t}$& --
& $\tanh(mt)$& --
& $\displaystyle\frac{1-e^{-\alpha t}}{1+\beta e^{-\alpha t}}$
& --
 \\
\midrule 
$L^2(x)$   &1.67& $1.35\times10^{1}$   & $1.14\times10^{1}$ & $15.00\%$ & $1.37\times10^{1}$ & $-1.72\%$ & $1.37$ & \highlightgreen{$89.80\%$} \\
$L^2(y)$  & 1.65& $1.13\times10^{1}$   & $1.01\times10^{1}$ & $10.67\%$ & $9.59$ & $15.29\%$ & $1.15$ & \highlightgreen{$89.88\%$} \\
$L^2(p_x)$& 1.25 & $1.14\times10^{1}$   & $9.66$ & $15.29\%$ & $1.18\times10^{1}$ & $-3.24\%$ & $1.18$ & \highlightgreen{$89.67\%$} \\
$L^2(p_y)$ &1.11 & $1.01\times10^{1}$   & $9.16$ & $8.96\%$ & $8.43$ & $16.21\%$ & $9.34\times10^{-1}$ & \highlightgreen{$90.71\%$} \\
\midrule
MSE$(x)$  &$2.79\times10^{-3}$ & $1.81\times10^{-1}$ & $1.31\times10^{-1}$ & $27.75\%$ & $1.87\times10^{-1}$ & $-3.48\%$ & $1.88\times10^{-3}$ & \highlightgreen{$98.96\%$} \\
MSE$(y)$  & $2.74\times10^{-3}$& $1.28\times10^{-1}$ & $1.02\times10^{-1}$ & $20.19\%$ & $9.19\times10^{-2}$ & $28.24\%$ & $1.31\times10^{-3}$ & \highlightgreen{$98.98\%$} \\
MSE$(p_x)$ & $1.57\times10^{-3}$& $1.30\times10^{-1}$ & $9.34\times10^{-2}$ & $28.24\%$ & $1.39\times10^{-1}$ & $-6.61\%$ & $1.39\times10^{-3}$ & \highlightgreen{$98.93\%$} \\
MSE$(p_y)$ & $1.23\times10^{-3}$& $1.01\times10^{-1}$ & $8.38\times10^{-2}$ & $17.12\%$ & $7.10\times10^{-2}$ & $29.80\%$ & $8.72\times10^{-4}$ & \highlightgreen{$99.14\%$} \\
\bottomrule 
\end{tabular}
\caption{
Error comparison between the ATLAS-NN transfer learning models for Hénon--Heiles system. Percentage improvements indicate relative error reduction with respect to the HNN baseline.
Cells shaded in light celadon green indicate improvements exceeding $80\%$.
For the ATLAS-NN models, the time-adaptive functions  are $f(t)=\tanh(mt)$ with $m=0.4699$, and
$f(t)=\frac{1-e^{-\alpha t}}{1+\beta e^{-\alpha t}}$ with $\alpha=0.5658$ and $\beta=1.1375$.
}
\label{tab:HH_transfer_errors}
\end{table}

The numerical results for the H\'{e}non--Heiles system, summarized in Table \ref{tab:HH_transfer_errors}, shows a significant disparity in predictive accuracy between the baseline and transfer-learned models. The transition from the \textit{HNN (Baseline)} to the \textit{HNN (Transfer)} yields consistent but modest improvements, with errors generally reduced by approximately $10\%$ to $25\%$. While this suggests that weight initialization from the source task provides a superior starting point, the absolute error magnitudes for both standard HNN configurations remain on the order of $10^{-1}$ to $10^{1}$, indicating an inability to fully capture the system's long-term behavior.
The \textit{ATLAS-NN} models, particularly the \textit{ATLAS-NN (exp)}, results in a dramatic improvement in performance. As highlighted by the shaded cells in Table  \ref{tab:HH_transfer_errors}, the \textit{transfer ATLAS-NN (exp)} model achieves improvements exceeding $89\%$ across all $L^{2}$ metrics. Specifically, the MSE for the coordinate variables $x$ and $y$ are reduced by two orders of magnitude compared to the baseline, dropping from roughly $10^{-1}$ to $10^{-3}$. The momentum error for $p_y$ shows the most significant reduction, with a final MSE of $8.72 \times 10^{-4}$ that is a reduction of nearly two orders of magnitude over the baseline HNN.
In contrast, the performance of the \textit{transfer ATLAS-NN (tanh)} is more nuanced. While it successfully reduces the MSE of $y$ and $p_y$ by nearly one-third compared to the transfer HNN, it occasionally exhibits slight performance degradation in other state variables. This suggests that the exponential time-adaptive function $f(t) = \frac{1 - e^{-\alpha t}}{1 + \beta e^{-\alpha t}}$ provides a more flexible and robust inductive bias for the complex, nonlinear oscillations of the H\'{e}non--Heiles system. 

\begin{figure}[H]
    \centering
    \includegraphics[width=1\linewidth]{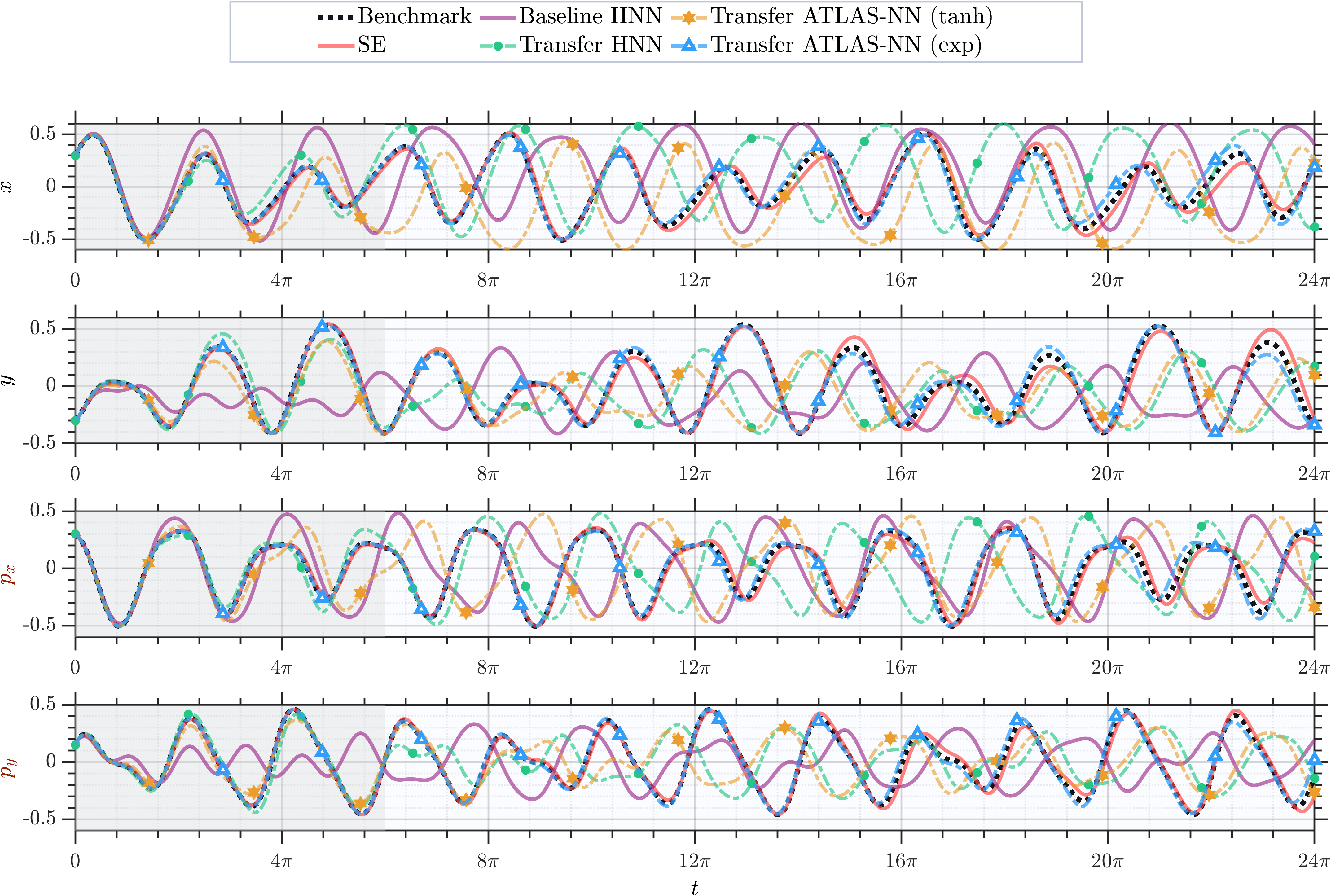}
    \caption{
 Comparison of long-term phase-space trajectories and temporal state evolutions for the H\'{e}non--Heiles system over the target interval $[0, 24\pi]$. The shaded region $[0, 6\pi]$ indicates the source interval used during the first task of the transfer learning process.
}
    \label{fig:HH_tran_error}
\end{figure}

\begin{figure}[H]
    \centering

    \begin{subfigure}[t]{\linewidth}
        \centering
        \includegraphics[width=\linewidth]{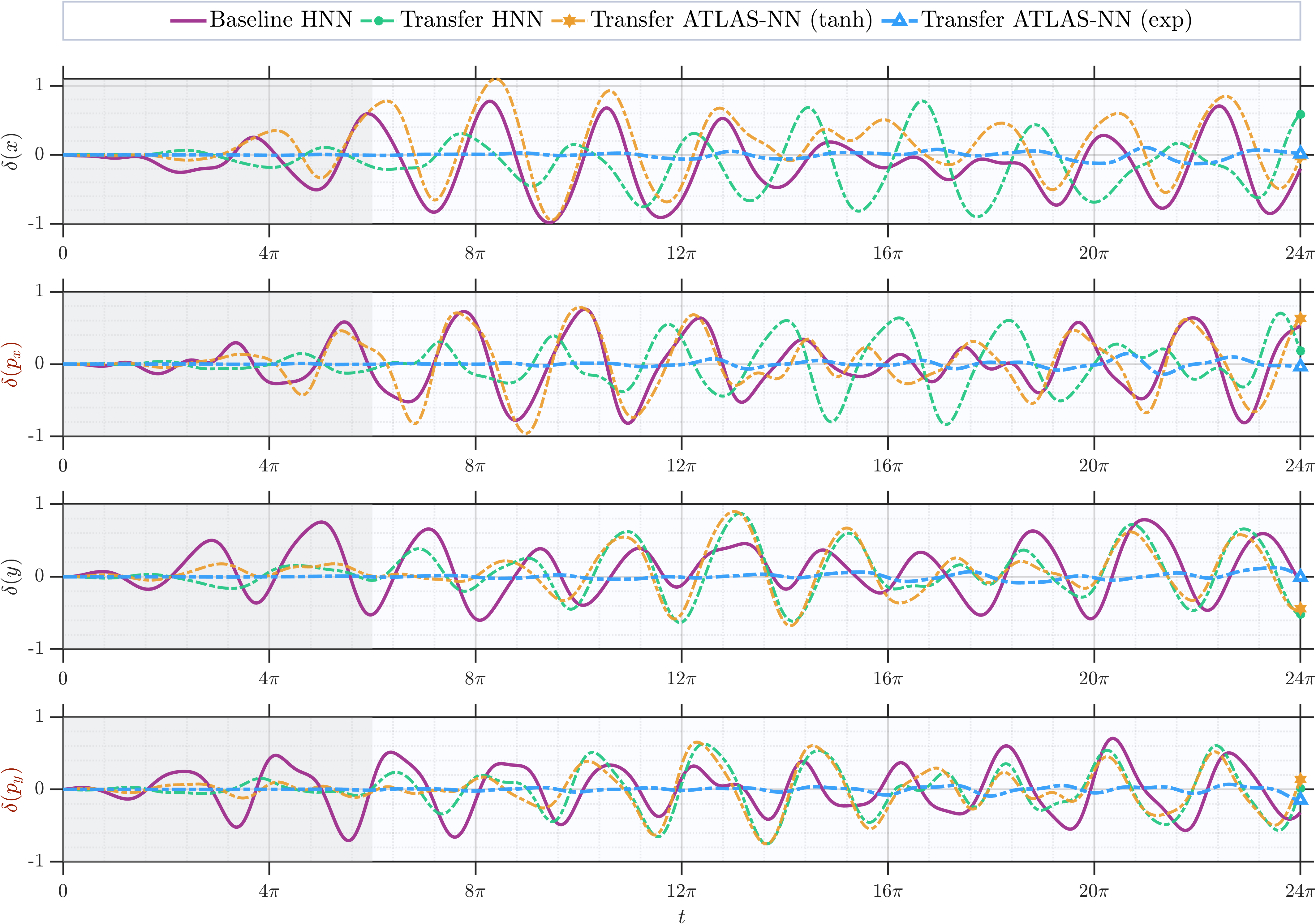}
        \caption{Error comparison}
        \label{fig:HH_trans_error}
    \end{subfigure}

    \vspace{0.6em}

    \begin{subfigure}[t]{\linewidth}
        \centering
        \includegraphics[width=\linewidth]{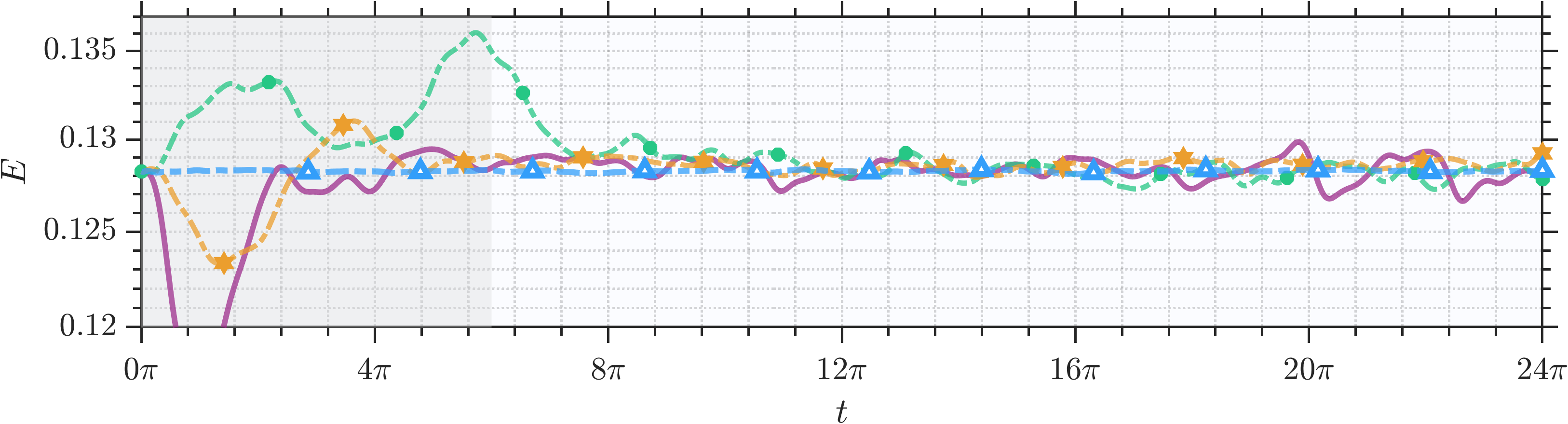}
        \caption{Energy evolution}
        \label{fig:HH_trans_energy}
    \end{subfigure}

    \caption{Time evolution of error and energy of the Hénon--Heiles system for different models in the target task $[0,24\pi]$. The left subpanels of Panel (a) show the phase space trajectories in the $(x,y)$, $(x,p_x)$, and $(y,p_y)$ planes, and the four right subpanels of Panel (a) illustrate the time evolution for each state variable: $x$, $y$, $p_x$, and $p_y$; Panel (b) shows the time evolution of the energy for the H\'enon-Heiles system..}
    \label{fig:HH_trans_ErrorCompare}
\end{figure}

The long-term predictive capabilities of the transfer learning frameworks are visually shown in Figure \ref{fig:HH_tran_error}, which describes the temporal evolution of the state variables ($x, y, p_x, p_y$) over the extended target interval $[0, 24\pi]$. The shaded region $[0, 6\pi]$ corresponds to the source interval, where all models—excluding the \textit{baseline HNN}—demonstrate initial synchronization. However, as the system evolves into the unshaded target regime, a clear divergence in performance becomes apparent.
Consistent with the $L^2$ and MSE metrics reported in Table \ref{tab:HH_transfer_errors}, the \textit{baseline HNN} (purple) and \textit{transfer HNN} (green) suffer from rapid phase drift and amplitude decay shortly after the source task time domain. These models fail to capture the nonlinear oscillations of the H\'{e}non--Heiles system, with their trajectories quickly becoming out of phase with the \textit{benchmark} solution.
In contrast, the \textit{ATLAS-NN} models show remarkable robustness. The \textit{transfer ATLAS-NN (exp)} (blue) maintains near-perfect consistency with the benchmark across the entire $24\pi$ duration, effectively mitigating the cumulative errors that plague non-adaptive models. The \textit{transfer ATLAS-NN (tanh)} (orange) also appears to more accurate compared to the standard HNNs, though it shows a slight increase in amplitude deviation toward the end of the time. 
The robustness of the transfer learning frameworks is further confirmed by the error and energy profiles illustrated in Figure \ref{fig:HH_trans_ErrorCompare}. As shown in the temporal error evolution (Figure \ref{fig:HH_trans_error}), the \textit{baseline HNN} (purple) and \textit{transfer HNN} (green) shows rapid, large-amplitude oscillations in $\delta x$, $\delta y$, and their respective momenta immediately following the transition from the source to the target regime at $t=6\pi$. This error growth correlates with the phase drift observed in the previous trajectory analysis, indicating the inability of standard models to maintain point-wise accuracy over long time intervals.
In contrast, the \textit{ATLAS-NN} models show improved accuracy. The \textit{transfer ATLAS-NN (exp)} (blue) maintains a remarkably flat error profile, with deviations remaining near zero across the entire $24\pi$ duration. While the \textit{transfer ATLAS-NN (tanh)} (orange) exhibits some oscillatory growth in the target region, its error magnitude remains significantly lower than that of the non-adaptive HNN models, demonstrating the efficacy of the transferred temporal bias.

Figure \ref{fig:HH_trans_energy} shows the energy evolution of the system for different models. The \textbf{baseline HNN} and \textbf{transfer HNN} show significant energy fluctuations and drift, particularly during the first half of the time domain. Notably, even the \textbf{transfer ATLAS-NN (tanh)} shows an initial dip in energy before stabilizing. Conversely, the \textbf{transfer ATLAS-NN (exp)} exhibits nearly perfect energy conservation, maintaining the Hamiltonian constant throughout the entire target task. These results indicate that the exponential time-adaptive framework not only preserves the geometric properties of the phase space but also ensures the physical consistency of the system's total energy, even when target task is far beyond the original source task.

\section{Conclusion and Future Work \label{sec:conclusion}}

In this work, we introduced the \textit{Adaptive Transfer Learnable Symplectic-aware Neural Network} (ATLAS-NN), a novel framework designed to overcome the structural limitations of standard Hamiltonian Neural Networks (HNNs) in long-time integration. By augmenting the HNN architecture with a learnable temporal scaling mechanism, ATLAS-NN effectively decouples the training time interval from the physical timescale of the dynamics. Our methodology utilizes a two-stage transfer learning strategy: a source task is first conducted over a short-time interval to identify the optimal temporal reparameterization, which is subsequently frozen and transferred to an extended target interval for fine-tuning the Hamiltonian representation.
Numerical experiments on the nonlinear oscillator and the H\'{e}non--Heiles system show that ATLAS-NN significantly outperforms both standard HNNs and traditional symplectic integrators. The results indicate that the learnable temporal parameters ($\alpha, \beta, m$) provide a robust inductive bias that stabilizes gradient flow and mitigates the accumulated phase errors typical of fixed-structure neural solvers. Specifically, for the H\'{e}non--Heiles system, the exponential ATLAS-NN variant achieved a reduction in prediction error by nearly two orders of magnitude compared to baseline models. Furthermore, the framework exhibits remarkable energy stability, maintaining the system's Hamiltonian nearly constant over integration periods exceeding the original training window by a factor of four.

Future research will focus on extending the ATLAS-NN framework to high-dimensional Hamiltonian systems and partial differential equations where multiscale features are even more complex \cite{mou2023efficient,majda2016introduction,qi2020using}. We also aim to investigate the integration of this adaptive mechanism into more general complex system such as Navier-Stokes equation and Boussinesq equations to enable the discovery of complex dynamics from noisy observational data \cite{chen2024stochastic,zhang2022convergence,zhang2021fast}. Finally, exploring dynamic 'thawing' strategies for the temporal kernel or parameterized propagator during the target task may provide additional flexibility for neural operators \cite{mou2025pas,mi2025pip} navigating significant non-stationary transitions over ultra-long temporal horizons.

\section*{Acknowledgements}
D.X is grateful to acknowledge the support of the National Natural Science Foundation of China (NSFC) under Grants No.12371428, and No. 11871435.
Y.Z is grateful to acknowledge the support of the National Natural Science Foundation of China (NSFC) under Grants No. 12401562 , No. 12571459, and No. 12241103.
\section*{Data and Code Availability}
The source code and associated datasets used in this study are publicly available at
\url{https://github.com/Zuoxy821/ATLAS-NN}.

\section*{Conflict of Interest Statement}
The authors declare that there are no conflicts of interest relevant to this manuscript.
No funding sources had any role in the study design, data collection, analysis,
interpretation, or the decision to submit this work for publication.

\bibliographystyle{unsrt}

\bibliography{ref}

@article{rackauckas2020universal,
  title={Universal Differential Equations for Scientific Machine Learning},
  author={Rackauckas, Christopher and Ma, Yingbo and Martensen, Julius and Warner, Collin and Zubov, Kirill and Supekar, Rohit and Skinner, Dominic and Ramadhan, Ali Jasim},
  journal={CoRR},
  year={2020}
}

@article{cai2021physics,
  title={Physics-informed neural networks (PINNs) for fluid mechanics: A review},
  author={Cai, Shengze and Mao, Zhiping and Wang, Zhicheng and Yin, Minglang and Karniadakis, George Em},
  journal={Acta Mechanica Sinica},
  volume={37},
  number={12},
  pages={1727--1738},
  year={2021},
  publisher={Springer}
}

@article{chen2018neural,
  title={Neural ordinary differential equations},
  author={Chen, Ricky TQ and Rubanova, Yulia and Bettencourt, Jesse and Duvenaud, David K},
  journal={Advances in neural information processing systems},
  volume={31},
  year={2018}
}

@inproceedings{lifourier,
  title={Fourier Neural Operator for Parametric Partial Differential Equations},
  author={Li, Zongyi and Kovachki, Nikola Borislavov and Azizzadenesheli, Kamyar and Bhattacharya, Kaushik and Stuart, Andrew and Anandkumar, Anima and others},
  booktitle={International Conference on Learning Representations}
}

@article{lu2021learning,
  title={Learning nonlinear operators via DeepONet based on the universal approximation theorem of operators},
  author={Lu, Lu and Jin, Pengzhan and Pang, Guofei and Zhang, Zhongqiang and Karniadakis, George Em},
  journal={Nature machine intelligence},
  volume={3},
  number={3},
  pages={218--229},
  year={2021},
  publisher={Nature Publishing Group UK London}
}

@article{thiyagalingam2022scientific,
  title={Scientific machine learning benchmarks},
  author={Thiyagalingam, Jeyan and Shankar, Mallikarjun and Fox, Geoffrey and Hey, Tony},
  journal={Nature Reviews Physics},
  volume={4},
  number={6},
  pages={413--420},
  year={2022},
  publisher={Nature Publishing Group UK London}
}

@article{carleo2019machine,
  title={Machine learning and the physical sciences},
  author={Carleo, Giuseppe and Cirac, Ignacio and Cranmer, Kyle and Daudet, Laurent and Schuld, Maria and Tishby, Naftali and Vogt-Maranto, Leslie and Zdeborov{\'a}, Lenka},
  journal={Reviews of Modern Physics},
  volume={91},
  number={4},
  pages={045002},
  year={2019},
  publisher={APS}
}

@article{donnelly2005symplectic,
  title={Symplectic integrators: An introduction},
  author={Donnelly, Denis and Rogers, Edwin},
  journal={American Journal of Physics},
  volume={73},
  number={10},
  pages={938--945},
  year={2005},
  publisher={AIP Publishing}
}

@article{bottasso1997new,
  title={A new look at finite elements in time: a variational interpretation of Runge-Kutta methods},
  author={Bottasso, Carlo L},
  journal={Applied Numerical Mathematics},
  volume={25},
  number={4},
  pages={355--368},
  year={1997},
  publisher={Elsevier Science Publishers BV Amsterdam, The Netherlands, The Netherlands}
}

@article{koopman1931hamiltonian,
  title={Hamiltonian systems and transformation in Hilbert space},
  author={Koopman, Bernard O},
  journal={Proceedings of the National Academy of Sciences},
  volume={17},
  number={5},
  pages={315--318},
  year={1931}
}

@book{goldstein2011classical,
  title={Classical mechanics},
  author={Goldstein, Herbert},
  year={2011},
  publisher={Pearson Education India}
}

@article{mattheakis2022hamiltonian,
  title={Hamiltonian neural networks for solving equations of motion},
  author={Mattheakis, Marios and Sondak, David and Dogra, Akshunna S and Protopapas, Pavlos},
  journal={Physical Review E},
  volume={105},
  number={6},
  pages={065305},
  year={2022},
  publisher={APS}
}

@book{wiggins2003introduction,
  title={Introduction to applied nonlinear dynamical systems and chaos},
  author={Wiggins, Stephen},
  year={2003},
  publisher={Springer}
}

@book{sakurai2020modern,
  title={Modern quantum mechanics},
  author={Sakurai, Jun John and Napolitano, Jim},
  year={2020},
  publisher={Cambridge University Press}
}

@book{goldstein1950classical,
  title={Classical mechanics},
  author={Goldstein, Herbert and Poole, Charles P and Safko, John},
  volume={2},
  year={1950},
  publisher={Addison-wesley Reading, MA}
}

@book{reichl2016modern,
  title={A modern course in statistical physics},
  author={Reichl, Linda E},
  year={2016},
  publisher={John Wiley \& Sons}
}

@book{zagoskin1998quantum,
  title={Quantum theory of many-body systems},
  author={Zagoskin, Alexandre M},
  volume={174},
  year={1998},
  publisher={Springer}
}

@article{greydanus2019hamiltonian,
  title={Hamiltonian neural networks},
  author={Greydanus, Samuel and Dzamba, Misko and Yosinski, Jason},
  journal={Advances in neural information processing systems},
  volume={32},
  year={2019}
}

@article{raissi2019physics,
  title={Physics-informed neural networks: A deep learning framework for solving forward and inverse problems involving nonlinear partial differential equations},
  author={Raissi, Maziar and Perdikaris, Paris and Karniadakis, George E},
  journal={Journal of Computational physics},
  volume={378},
  pages={686--707},
  year={2019},
  publisher={Elsevier}
}

@book{abraham1978foundations,
  title     = {Foundations of Mechanics},
  author    = {Abraham, Ralph and Marsden, Jerrold E.},
  edition   = {2},
  publisher = {Benjamin/Cummings},
  year      = {1978},
  isbn      = {9780805301025}
}

@book{marsden2013introduction,
  title={Introduction to mechanics and symmetry: a basic exposition of classical mechanical systems},
  author={Marsden, Jerrold E and Ratiu, Tudor S},
  volume={17},
  year={2013},
  publisher={Springer Science \& Business Media}
}

@book{leimkuhler2004simulating,
  title={Simulating hamiltonian dynamics},
  author={Leimkuhler, Benedict and Reich, Sebastian},
  number={14},
  year={2004},
  publisher={Cambridge university press}
}

@article{hairer2006structure,
  title={Structure-preserving algorithms for ordinary differential equations},
  author={Hairer, Ernst and Lubich, Christian and Wanner, Gerhard},
  journal={Geometric numerical integration},
  volume={31},
  year={2006},
  publisher={Springer-Verlag Berlin}
}

@article{desai2021port,
  title={Port-Hamiltonian neural networks for learning explicit time-dependent dynamical systems},
  author={Desai, Shaan A and Mattheakis, Marios and Sondak, David and Protopapas, Pavlos and Roberts, Stephen J},
  journal={Physical Review E},
  volume={104},
  number={3},
  pages={034312},
  year={2021},
  publisher={APS}
}

@article{pellegrin2022transfer,
  title={Transfer learning with physics-informed neural networks for efficient simulation of branched flows},
  author={Pellegrin, Rapha{\"e}l and Bullwinkel, Blake and Mattheakis, Marios and Protopapas, Pavlos},
  journal={arXiv preprint arXiv:2211.00214},
  year={2022}
}

@article{mou2025pas,
  title={PAS-Net: Physics-informed Adaptive Scale Deep Operator Network},
  author={Mou, Changhong and Zhang, Yeyu and Zhu, Xuewen and Zhuang, Qiao},
  journal={arXiv preprint arXiv:2511.14925},
  year={2025}
}

@article{mi2025pip,
  title={PIP$^2$ Net: Physics-informed Partition Penalty Deep Operator Network},
  author={Mi, Hongjin and Lun, Huiqiang and Mou, Changhong and Zhang, Yeyu},
  journal={arXiv preprint arXiv:2512.15086},
  year={2025}
}

@article{chen2024stochastic,
  title={A stochastic precipitating quasi-geostrophic model},
  author={Chen, Nan and Mou, Changhong and Smith, Leslie M and Zhang, Yeyu},
  journal={Physics of Fluids},
  volume={36},
  number={11},
  year={2024},
  publisher={AIP Publishing}
}

@article{zhang2022convergence,
  title={Convergence to precipitating quasi-geostrophic equations with phase changes: asymptotics and numerical assessment},
  author={Zhang, Yeyu and Smith, Leslie M and Stechmann, Samuel N},
  journal={Philosophical Transactions of the Royal Society A: Mathematical, Physical and Engineering Sciences},
  volume={380},
  number={2226},
  year={2022},
  publisher={The Royal Society}
}

@article{mou2023combining,
  title={Combining stochastic parameterized reduced-order models with machine learning for data assimilation and uncertainty quantification with partial observations},
  author={Mou, Changhong and Smith, Leslie M and Chen, Nan},
  journal={Journal of Advances in Modeling Earth Systems},
  volume={15},
  number={10},
  pages={e2022MS003597},
  year={2023},
  publisher={Wiley Online Library}
}

@article{weiss2016survey,
  title={A survey of transfer learning},
  author={Weiss, Karl and Khoshgoftaar, Taghi M and Wang, DingDing},
  journal={Journal of Big data},
  volume={3},
  number={1},
  pages={9},
  year={2016},
  publisher={Springer}
}

@incollection{torrey2010transfer,
  title={Transfer learning},
  author={Torrey, Lisa and Shavlik, Jude},
  booktitle={Handbook of research on machine learning applications and trends: algorithms, methods, and techniques},
  pages={242--264},
  year={2010},
  publisher={IGI Global Scientific Publishing}
}

@article{bender1969anharmonic,
  title={Anharmonic oscillator},
  author={Bender, Carl M and Wu, Tai Tsun},
  journal={Physical Review},
  volume={184},
  number={5},
  pages={1231},
  year={1969},
  publisher={APS}
}

@article{sharma2022hamiltonian,
  title={Hamiltonian operator inference: Physics-preserving learning of reduced-order models for canonical Hamiltonian systems},
  author={Sharma, Harsh and Wang, Zhu and Kramer, Boris},
  journal={Physica D: Nonlinear Phenomena},
  volume={431},
  pages={133122},
  year={2022},
  publisher={Elsevier}
}

@article{ruiz2023neural,
  title={Neural ODE control for classification, approximation, and transport},
  author={Ruiz-Balet, Domenec and Zuazua, Enrique},
  journal={SIAM Review},
  volume={65},
  number={3},
  pages={735--773},
  year={2023},
  publisher={SIAM}
}

@inproceedings{feng2006symplectic,
  title={The symplectic methods for the computation of Hamiltonian equations},
  author={Feng, Kang and Qin, Meng-zhao},
  booktitle={Numerical Methods for Partial Differential Equations: Proceedings of a Conference held in Shanghai, PR China, March 25--29, 1987},
  pages={1--37},
  year={2006},
  organization={Springer}
}

@article{zhuang2020comprehensive,
  title={A comprehensive survey on transfer learning},
  author={Zhuang, Fuzhen and Qi, Zhiyuan and Duan, Keyu and Xi, Dongbo and Zhu, Yongchun and Zhu, Hengshu and Xiong, Hui and He, Qing},
  journal={Proceedings of the IEEE},
  volume={109},
  number={1},
  pages={43--76},
  year={2020},
  publisher={Ieee}
}

@article{mou2023efficient,
  title={An efficient data-driven multiscale stochastic reduced order modeling framework for complex systems},
  author={Mou, Changhong and Chen, Nan and Iliescu, Traian},
  journal={Journal of Computational Physics},
  volume={493},
  pages={112450},
  year={2023},
  publisher={Elsevier}
}

@book{majda2016introduction,
  title={Introduction to turbulent dynamical systems in complex systems},
  author={Majda, Andrew J},
  year={2016},
  publisher={Springer}
}

@article{qi2020using,
  title={Using machine learning to predict extreme events in complex systems},
  author={Qi, Di and Majda, Andrew J},
  journal={Proceedings of the National Academy of Sciences},
  volume={117},
  number={1},
  pages={52--59},
  year={2020},
  publisher={National Academy of Sciences}
}

@article{zhang2021fast,
  title={Fast-wave averaging with phase changes: asymptotics and application to moist atmospheric dynamics},
  author={Zhang, Yeyu and Smith, Leslie M and Stechmann, Samuel N},
  journal={Journal of Nonlinear Science},
  volume={31},
  number={2},
  pages={38},
  year={2021},
  publisher={Springer}
}
\end{document}